%% file: main.tex
\documentclass[sigconf,nonacm]{acmart}
\usepackage{listings}
\usepackage{xcolor}
\usepackage{setspace}
\usepackage{graphicx}
\usepackage{subcaption}
\usepackage{ragged2e}
\usepackage{float}
\usepackage{tabularx} 
\usepackage{xurl}
\usepackage{hyperref}
\usepackage{tcolorbox}

\lstdefinelanguage{json}{
  basicstyle=\ttfamily\small\setstretch{1.2}, 
  showstringspaces=false,
  breaklines=true,
  frame=none, 
  numbers=none, 
  backgroundcolor=\color{white}, 
  literate=
   *{0}{{{\color{blue}0}}}{1}
    {1}{{{\color{blue}1}}}{1}
    {2}{{{\color{blue}2}}}{1}
    {3}{{{\color{blue}3}}}{1}
    {4}{{{\color{blue}4}}}{1}
    {5}{{{\color{blue}5}}}{1}
    {6}{{{\color{blue}6}}}{1}
    {7}{{{\color{blue}7}}}{1}
    {8}{{{\color{blue}8}}}{1}
    {9}{{{\color{blue}9}}}{1}
    {:}{{{\color{black}:}}}{1}
    {,}{{{\color{black},}}}{1}
    {"}{{{\color{orange}"}}}{1}
}

\AtBeginDocument{%
  }

\renewcommand\footnotetextcopyrightpermission[1]{}
\copyrightyear{2026}
\acmDOI{XXXXXXX.XXXXXXX}
\acmConference[Conference acronym 'XX]{Make sure to enter the correct
  conference title from your rights confirmation email}{June 03--05,
  2018}{Woodstock, NY}

\lstdefinestyle{json}{
    basicstyle=\ttfamily,
    numbers=left,
    numberstyle=\tiny\color{gray},
    stepnumber=1,
    numbersep=8pt,
    showstringspaces=false,
    breaklines=true,
    frame=lines,
    backgroundcolor=\color{lightgray!20},
    keywordstyle=\color{blue},
    stringstyle=\color{red},
    commentstyle=\color{green!60!black},
    morekeywords={true,false,null}
}
\begin{document}

\title[AwareLLM: A Proactive Multimodal Human-AI Collaborative Ecosystem]{AwareLLM: A Proactive Multimodal Ecosystem for Personalized Human-AI Collaboration to Enhance Productivity}

\author{Amog Rao}
\orcid{0009-0000-4877-6834}
\authornote{These authors contributed equally to this work.}
\affiliation{%
  \institution{Microsoft Research}
  \city{Bangalore}
  \country{India}
}
\email{t-amograo@microsoft.com}

\author{Utkarsh Agarwal}
\orcid{0009-0009-4055-9902}
\authornotemark[1]
\affiliation{%
  \institution{HTI Lab, Plaksha University}
  \city{Mohali}
  \country{India}
}
\email{utkarsh.agarwal@plaksha.edu.in}

\author{Amol Harsh}
\orcid{0009-0007-9529-6850}
\affiliation{%
  \institution{Mohamed Bin Zayed University of Artificial Intelligence}
  \city{Abu Dhabi}
  \country{United Arab Emirates}
}
\email{amol.harsh@mbzuai.ac.ae}

\author{Siddharth Siddharth}
\orcid{0000-0002-1001-8218}
\affiliation{%
  \institution{HTI Lab, Plaksha University}
  \city{Mohali}
  \country{India}
}
\email{siddharth.s@plaksha.edu.in}
\renewcommand{\shortauthors}{Rao et al.}

\begin{abstract}
Information workers' productivity is significantly influenced by their cognitive states and physiological responses. AI assistants such as ChatGPT, Copilot, and others have become integral components of knowledge-intensive workplaces. These AI assistants utilize pre-defined user preferences and chat interaction histories, thus confining themselves to reactive exchanges, lacking sufficient adaptability. Consequently, they fail to cater to individual user preferences and are unable to adapt to their psychophysiological states, diminishing potential productivity gains. To bridge this gap, we introduce AwareLLM, a novel multimodal framework that integrates egocentric vision, pupillometry, eye-gaze tracking, posture detection, cardiac activity, and the inferencing capabilities of large language models (LLMs) to create a proactive and context-aware ecosystem. AwareLLM dynamically adapts to users' psychophysiological states while analyzing temporal patterns and behavioral tendencies to provide personalized and timely interventions. We evaluated AwareLLM through a user study with 20 participants, comparing it to a standard LLM assistant across multiple tasks. Our results show statistically significant improvements in task performance, along with reductions in cognitive fatigue and mental demand. Participants described AwareLLM’s personalized interventions as timely and relevant, helping them boost their confidence and deepen engagement with their work. AwareLLM opens new avenues for Human-AI collaboration where technology adapts to our needs rather than us adhering to technological constraints.
\end{abstract}




\begin{teaserfigure}
  \centering
  \includegraphics[width=\linewidth]{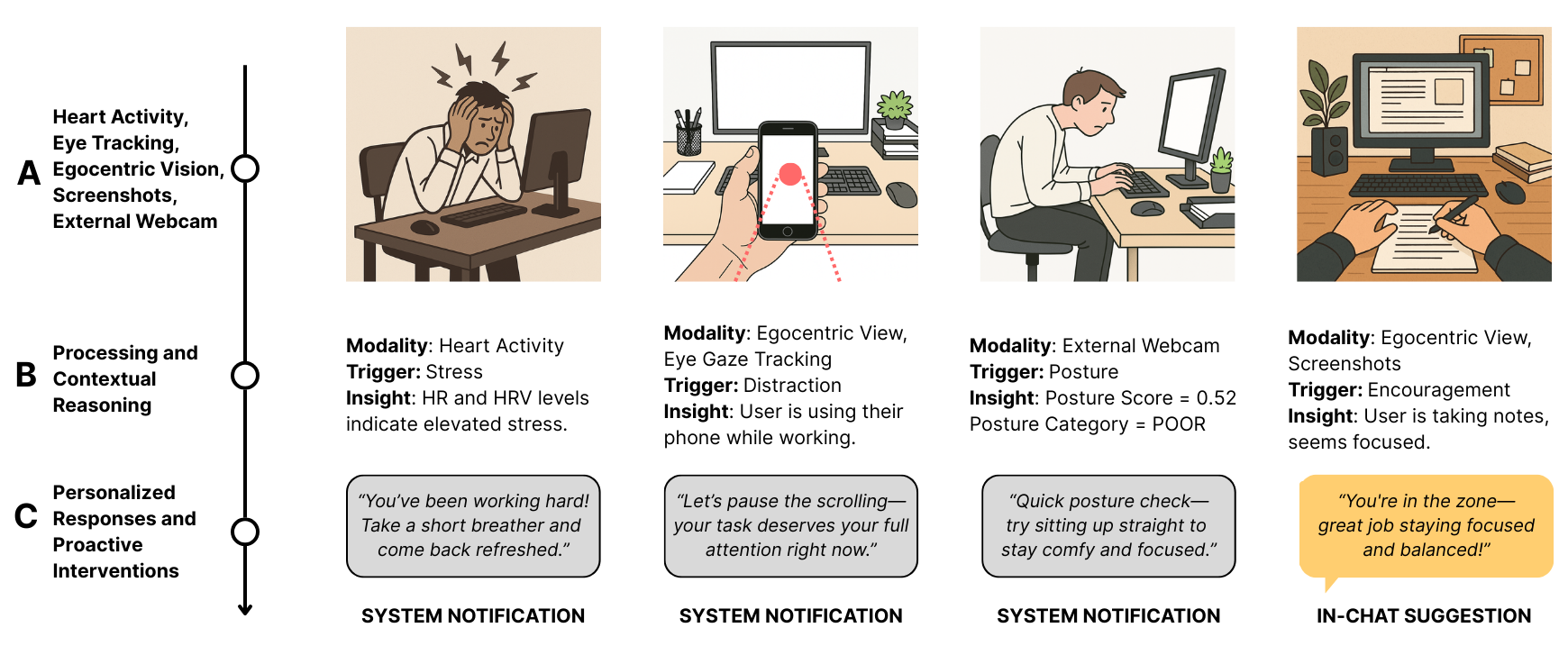}
  \caption{Introducing AwareLLM, a proactive, multimodal AI assistant designed to enhance productivity through personalized and adaptive support. It (A) continuously captures biosignals and contextual cues through a range of sensors, (B) interprets users’ cognitive and physical states and (C) boosts productivity by delivering timely, tailored interventions to help maintain focus, reduce strain, and support well-being throughout the workday.}
  \Description{A horizontal timeline illustrating the AwareLLM system across three stages. On the left, labeled 'A', multiple input sources are listed: biosensors, egocentric vision, screenshots, and external webcam. The middle section, labeled 'B', shows four illustrations representing different triggers: a stressed person (based on cardiac activity), someone using a phone while working (distraction), a person sitting slouched (poor posture), and a person taking notes at a computer (focused). Each trigger is associated with a specific modality and insight. On the right, labeled 'C', personalized responses are shown: system notifications for stress, distraction, and posture correction, and an in-chat encouragement for focused work.}
  \label{fig:teaser}
\end{teaserfigure}


\maketitle
{}

\section{Introduction}

Information workers’ productivity is significantly influenced by their cognitive states and physiological responses \cite{Alam2024, Shibaoka2023, Weintraub2023}. In today’s knowledge-intensive workplaces, where tasks such as writing reports, coding software, analyzing data, conducting literature reviews, drafting emails, and designing interfaces are routine. Artificial Intelligence (AI) assistants like ChatGPT, GitHub Copilot, and Claude have become indispensable. These systems excel at generating content, summarizing information, and supporting data interpretation \cite{Liu2025, 10.1145/3633453}.Their contextually relevant suggestions help alleviate users' cognitive load.

To address the limitations of reactive AI, recent research has explored physio-adaptive multimodal frameworks. For instance, concurrent systems like AdaptAI \cite{10.1145/3706599.3720284} integrate LLM-based workflows with physical movement (IMU) and audio cues to deliver personalized well-being interventions. However, these systems predominantly operate on slow temporal resolutions, such as aggregating data into 15-minute routine tables, and rely on macroscopic behavioral shifts rather than direct cognitive metrics. Consequently, they fail to capture the high-frequency micro-fluctuations in cognitive load, visual attention, and acute autonomic stress that characterize complex knowledge work. A disconnect persists between users' rapid, real time psychophysiological data and the slower, routine-based frameworks of these recent AI systems.

To address these limitations, we introduce \textbf{AwareLLM}: a system that reimagines workplace productivity by aligning AI assistance with human cognitive and physiological responses. AwareLLM continuously monitors biomarkers such as posture, heart rate variability, eye movements, and on-screen activity, processing them into interpretable measures of stress, attention, and posture. Leveraging a Large Language Model (LLM), it correlates these signals with the user’s task context to proactively deliver tailored recommendations, such as suggesting posture adjustments, mental breaks, or focused task decomposition. By aligning interventions with real time psychophysiological cues, AwareLLM aims to actively enhance productivity and well-being.

AwareLLM represents a paradigm shift in Human–AI collaboration, transitioning from reactive, one-way systems to proactive, context-sensitive assistants that adapt dynamically to users’ states and workflows. This two-way communication model preserves human agency while augmenting cognitive performance, fundamentally redefining collaboration in information-centric environments.

To validate the need for such a system and to understand the shortcomings of current AI models and agents, we conducted a formative study with 72 participants spanning researchers, developers, analysts, and students. Through an online survey combining structured and open-ended questions, we examined AI tool usage, software dependencies, and coping strategies for workplace challenges. Quantitative and thematic analyses of these responses informed the design and evaluation of AwareLLM.

To evaluate the performance of AwareLLM, we conducted a user study with 20 participants who engaged in three distinct activities: literature review, front-end web development, and data science, informed by our formative study. The control condition consisted of a standard LLM-based assistant that provided only reactive, prompt-driven support, lacking the proactive, context-aware interventions enabled by our AwareLLM ecosystem (intervention). The enhanced multimodal assistance offered by AwareLLM led to substantial improvements: NASA-TLX \cite{HART1988139} ratings showed that with AwareLLM, mental and temporal demands were reduced by over 22\% on average, while performance ratings improved by more than 15\% compared to the control condition. Furthermore, the post-study questionnaire reinforced these findings, with users reporting enhanced focus, improved work quality, and better time management when using AwareLLM. Complementing these subjective outcomes, expert evaluations of participants’ task outputs further confirmed significant improvements in task quality, completeness, and analytical rigor across all domains. Through context-aware suggestions, ranging from keyword recommendations and structural frameworks during literature reviews to tailored coding assistance and dynamic data visualizations, AwareLLM not only minimized cognitive overload but also elevated overall productivity.

Our research makes the following key contributions:
\begin{enumerate}
    \item \textbf{Proactive Biosensory Framework:} We propose a novel multimodal framework that seamlessly incorporates physiological and contextual signals into an LLM-driven workflow, moving from reactive chatbots to proactive, context-aware, and tone-adaptive assistance.
    \item \textbf{Formative Assessments:} A formative study involving 72 participants from diverse professional backgrounds uncovers detailed insights into their AI tool usage patterns, productivity hurdles, and personalized work strategies, which directly inform the design and refinement of our context-aware AI assistant system.
    
    \item \textbf{Generalizable Across Workflows:} By testing interventions across diverse cognitive workflows, spanning front-end web development, literature review, and data analysis, we demonstrate AwareLLM's dynamic adaptability and its potential as a universal augmentation layer for productivity and well-being in modern workplaces.
    
    \item \textbf{Real-World Impact and Validation:} Through a preliminary study with 20 participants, we reveal key limitations of reactive AI tools and show how AwareLLM boosts productivity, reduces cognitive fatigue, and enhances user satisfaction. This paves the way for a transformative future of human-centered AI in demanding work environments.

\end{enumerate}

By addressing both the ``mind'' (cognitive load) and ``body'' (physical posture, stress signals), AwareLLM brings a human-centered focus to AI-assisted productivity that extends beyond text-based conversation alone. The codebase for AwareLLM will be publicly released upon publication, enabling the community to explore, adapt, and build upon our work.

\section{Literature Survey}
Human-AI collaboration has rapidly evolved from early explorations of context-aware systems that harnessed basic environmental sensor data to enable user-centric interactions. As the field advanced, the incorporation of richer physiological signals paved the way for systems that could infer users’ internal states and adjust seamlessly in real time. Today, advancements in generative models and multimodal fusion are not only driving the development of intelligent, personalized interfaces but are also fostering a deeper, synergistic partnership between humans and AI.


\subsection{Foundational Ambient Computing}

Before 2015, the research community was deeply engaged in creating environments where technology blended seamlessly into everyday life as presented in Mark Weiser’s groundbreaking 1991 article, \textit{The Computer for the 21st Century} \cite{Weiser1991}. Weiser’s work introduced a practical framework for ubiquitous computing by conceptualizing devices in terms of “tabs,” “pads,” and “boards” that were meant to disappear into the background of everyday environments—exemplifying how technology could become ambient and non-intrusive. Meanwhile, pioneering context-aware systems, built on frameworks like Dey’s Context Toolkit, began to harness real time sensor data such as GPS-based location, timestamps, and user-defined settings to dynamically adjust interfaces and functionalities \cite{Dey2001}. At this stage, the integration of richer physiological or biosignal data was largely unexplored, laying a foundational framework that later inspired more complex approaches to adaptive, context-sensitive systems. 

\subsection{Physio-Adaptive and Proactive Interfaces}
Between 2015 and 2016, several important studies were conducted that shaped our understanding of how context and physiological signals can improve Human-AI collaboration. Initially, Mayer \textit{et al.} introduced a method called context-aware social matching. Unlike older computing systems that only cared about where people were (their location) while making socially intelligent decisions, this approach also considered who they know (relationships) and how they behave or feel (personal cues) \cite{Mayer2015}. 


Around the same time, Schmidt highlighted the potential of using biosignals (measurements from the body such as heart rate or skin conductance) as a new way to interact with computers, suggesting that these signals could help systems adjust automatically without explicit user commands \cite{Schmidt2016}. Later, Howell \textit{et al.} demonstrated that interpreting these body signals is not straightforward by using a special garment that changes color with skin conductance. Their work reminded us that such physiological data should be seen as social signals that provide clues about a person’s state, rather than as simple, direct indicators of emotions \cite{Howell2016}. Together, these studies brought attention back to the importance of context and bodily signals as key elements for creating smarter, more intuitive Human-AI collaborative systems.

These innovative ideas from the mid-2010s evolved into practical interfaces that bridged human behavior and computer interaction during 2016--2017. For example, the FlowLight study reduced workplace interruptions by automatically detecting when a person was busy—meaning the system inferred a “busy” state without the user having to manually set it \cite{Zuger2017}. At the same time, researchers worked on smarter digital notifications: one study by  Morrison \textit{et al.} explored how to time these alerts more effectively, while another by Zhang and Sundar focused on personalizing notifications to fit the user’s context \cite{Morrison2017,Zhang2019}. Additionally, advances in physiological sensing emerged as another prominent way to assess a user's mental state in a non-intrusive manner. For instance, the EEG-based \emph{AttentivU} headband detected lapses in attention and provided auditory feedback (sound alerts) to help refocus the user \cite{AttentivU2018}. These prototypes validated that biosignals
can meaningfully modulate interface behavior in real time, ushering in the era of
physio-adaptive HCI.


Between 2019 and 2021, adaptive systems evolved to not just react to users, but to anticipate their needs. For example, Lindlbauer \textit{et al.} showcased context-aware adaptation in mixed-reality heads-up displays that adjusted information density based on the user's estimated cognitive load. The system was able to predict when a user might be overwhelmed and simplified the display \cite{Lindlbauer2019}. Complementary advances in multimodal interaction techniques further improved assistive technologies by integrating inputs from different sources \cite{Karpov2014}. Rao \textit{et al.} developed a simulated marksmanship task in immersive virtual reality, using biosensors and behavioral signals to assess cognitive load and predict performance in dynamic environments \cite{Rao2020}. Furthermore, Meurisch \textit{et al.} found that for proactive assistants to be trusted, their timing and the relevance of their suggestions had to match the current situation perfectly \cite{Meurisch2020}. Recognizing the growing importance of collaboration between humans and AI, the NeurIPS 2020 Cooperative AI Workshop explored strategies for fostering cooperation not only among AI agents but also between humans and AI systems to improve joint performance and alignment. Together, these contributions shifted the focus from reactive adjustments to predictive and collaborative adaptations, highlighting the need for detailed user models that blend activity, workload, and affect.
\subsection{Generative AI and Multimodal Fusion}
After 2023, the emergence of LLMs and other generative AI models sparked a new era of context-aware systems that intelligently respond to both internal and external cues. For instance, Cao \textit{et al.} integrated generative models with adaptive user interfaces to enable real time adaptation across multiple data types, such as tabular datasets, textual inputs, and user-generated parameters \cite{Cao2025}. Chiossi \textit{et al.} demonstrated that adjusting visual complexity in virtual reality based on electrodermal activity, a measure of skin conductance linked to emotional arousal, can help boost working memory \cite{Chiossi2023}. Similarly, Kosch \textit{et al.} provided a comprehensive review of how to measure cognitive workload by combining information from multiple sources, known as multimodal fusion \cite{Kosch2023}. Meanwhile, Fan \textit{et al.}'s \emph{ContextCam} took a creative approach by linking environmental cues with AI-powered image generation to produce more relevant visuals \cite{Fan2024ContextCam}. These innovations collectively highlight a shift toward AI systems that are not only context-aware but also deeply personalized, laying the groundwork for the next generation of truly human-centric intelligent assistants.

In summary, across the past decade, the literature charts a clear progression: from
context recognition \cite{Mayer2015, Perdikis2021} and biosignal
exploration \cite{Schmidt2016,Howell2016} to physio-adaptive
interfaces \cite{Zuger2017,AttentivU2018}, proactive
modeling \cite{Lindlbauer2019,Meurisch2020}, and
collaborative frameworks \cite{Lai2022,Moge2022}. Recent
works pair multimodal sensing with generative AI
\cite{Fan2024ContextCam,Xu2024}, yet current systems often treat
physiological context as an \emph{auxiliary} feature rather than a
first-class driver of adaptation. For example, imagine a smart fitness app that occasionally checks your heart rate, rather than continuously using it to adjust your workout in real time. Moreover, proactivity policies
remain hand-crafted, lacking unified mechanisms to reason over rich
psychophysiological states, user workflows, and temporal work patterns.


Notably, recent concurrent work has explored personalized AI interventions via multimodal sensing. AdaptAI \cite{10.1145/3706599.3720284}, for example, combines egocentric vision, audio, and motion data to estimate user routines, employing an LLM to generate well-being interventions at 15-minute intervals. While effective for macroscopic routine management, AdaptAI lacks the capability to measure direct cognitive load and blocks interventions entirely during high-criticality tasks. AwareLLM advances this paradigm by replacing audio and motion tracking with direct cognitive sensing (pupillometry and eye-gaze tracking), and by introducing a dual-loop architecture that separates rapidly changing behavioral and contextual signals from slower physiological dynamics. Beyond sensing, AwareLLM performs cross-modal contextual reasoning across these signals and distinguishes task-focused in-chat assistance from user-focused system notifications, selecting the intervention type and delivery channel according to the inferred state, context, and urgency.

\textit{AwareLLM} therefore couples fine-grained user state estimation with the flexible reasoning of LLMs, enabling anticipatory, personalized interventions that align with the trajectory of human--AI collaboration illuminated by the past decade’s scholarship.

\section{Formative Study}

To validate the need for a context-aware AI assistant system and explore the scope of integrating psychophysiological states with LLMs, we conducted a precursor survey with diverse groups of information workers (including researchers, software developers, data analysts, and students). This section describes our methodology, findings, and how these insights informed the design and implementation of AwareLLM.

\subsection{Procedure}

We recruited 72 participants by distributing an online survey via university mailing lists and internal forums. The respondent pool was composed of knowledge workers in a university setting, including students (undergraduate and graduate), teaching assistants, administrative staff, PhD scholars, and research assistants. There were no specific exclusion criteria, and participation was voluntary and uncompensated. This recruitment strategy, while a convenience sample, provided a focused lens on individuals engaged in the types of knowledge-intensive tasks, such as coding, research, and data analysis, that AwareLLM is designed to support. The survey consisted of 10 questions designed to capture a mix of quantitative and qualitative data. The questions covered the following key areas:

\begin{enumerate}
    \item Demographic information and role
    \item Software tools primarily used
    \item AI tools frequently used and usage frequency
    \item Activities that occupy the largest portion of participants' day
    \item Purposes for which AI tools are primarily used
    \item Challenges that AI tools help overcome
    \item Methods used to enhance productivity when encountering roadblocks
    \item Perceived impact of AI tools on creativity and independent thinking
    \item Perception of AI tools' personalization to individual usage patterns
    \item Open-ended feedback on desired AI capabilities to capture bio-sensor feedback to tune responses according to users' needs
\end{enumerate}

The survey took an average of 5 minutes and 39 seconds to complete. We analyzed the structured questions quantitatively and performed a thematic analysis on the open-ended responses to identify recurring patterns and core needs.

\subsection{Findings}

We analyzed participants' responses using a mixed-methods approach, with quantitative analysis for the structured questions and thematic analysis for the open-ended question (Q10). This analysis provided key insights into AI usage patterns, productivity challenges, and intervention preferences, which informed the design of our experimental procedures.

\begin{figure}[htbp]
  \centering
  \includegraphics[width=0.8\linewidth]{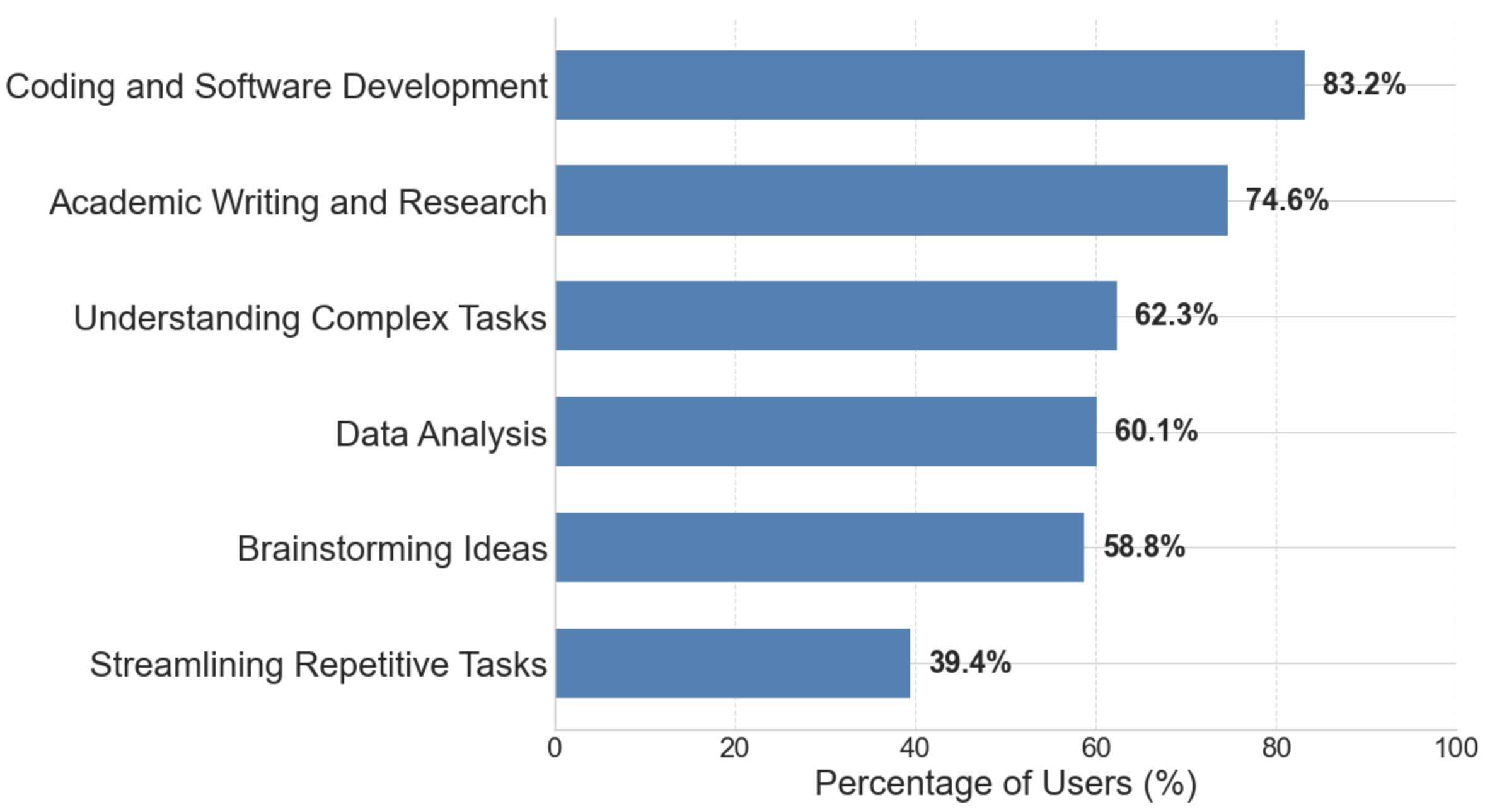}
  \caption{Motivations Behind Al Tool Adoption}
  \Description{A horizontal bar chart titled 'Motivations Behind AI Tool Adoption', showing the percentage of users who adopt AI tools for various purposes. The top motivation is 'Coding and Software Development' at 83.2\%, followed by 'Academic Writing and Research' at 74.6\%, 'Understanding Complex Tasks' at 62.3\%, 'Data Analysis' at 60.1\%, 'Brainstorming Ideas' at 58.8\%, and 'Streamlining Repetitive Tasks' at 39.4\%. The x-axis represents the percentage of users, ranging from 0 to 100.}
  \label{fig:form_1}
\end{figure}

\begin{figure}[htbp]
  \centering
  \includegraphics[width=0.8\linewidth]{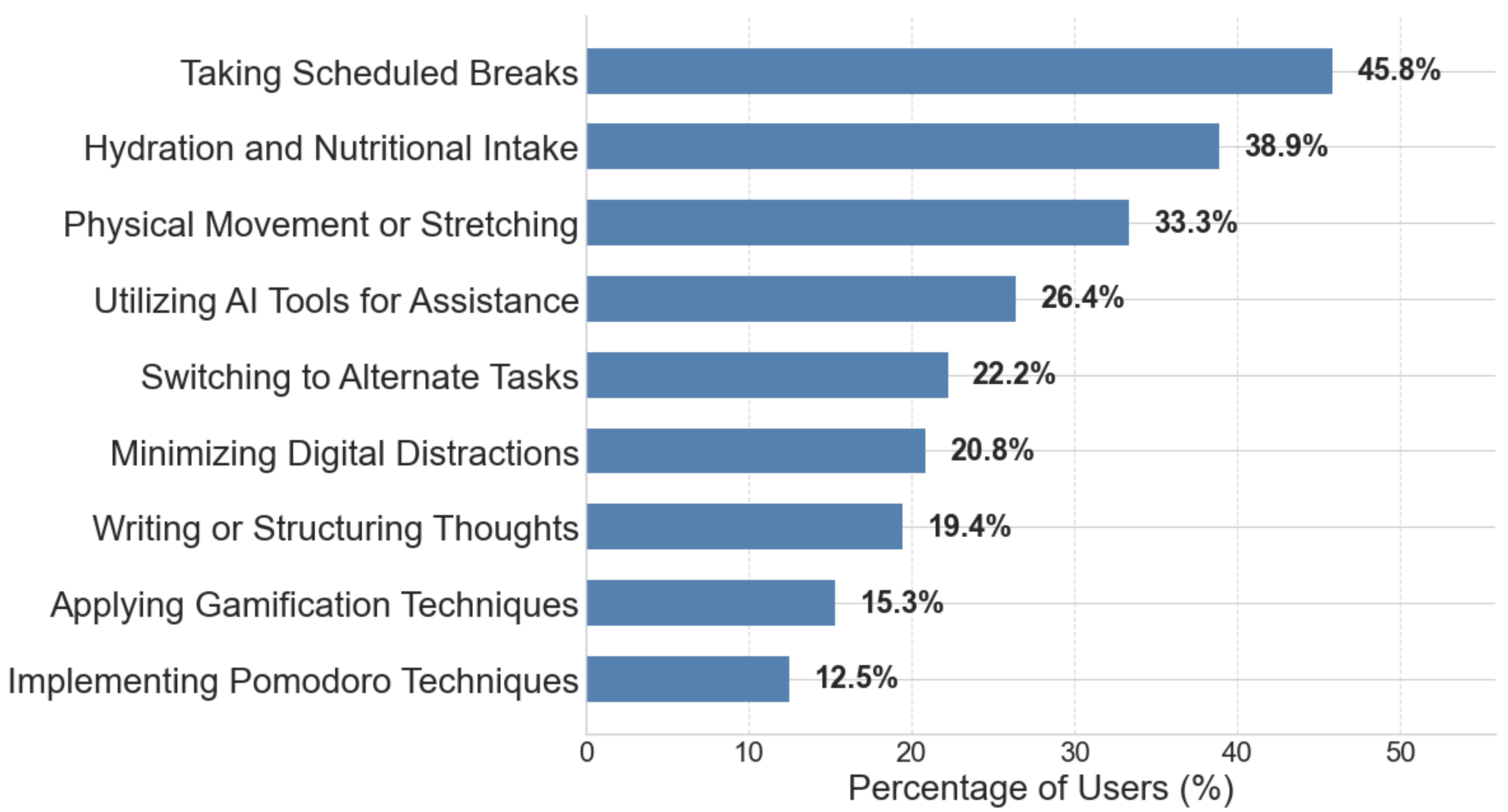}
  \caption{User strategies for managing obstacles that hinder their productivity}
  \Description{A horizontal bar chart titled 'User Strategies for Managing Productivity Obstacles', showing the percentage of users who use various strategies. The most common strategy is 'Taking Scheduled Breaks' at 45.8\%, followed by 'Hydration and Nutritional Intake' at 38.9\%, 'Physical Movement or Stretching' at 33.3\%, 'Utilizing AI Tools for Assistance' at 26.4\%, 'Switching to Alternate Tasks' at 22.2\%, 'Minimizing Digital Distractions' at 20.8\%, 'Writing or Structuring Thoughts' at 19.4\%, 'Applying Gamification Techniques' at 15.3\%, and 'Implementing Pomodoro Techniques' at 12.5\%. The x-axis represents percentage of users from 0 to 50.}
  \label{fig:form_2}
\end{figure}

\subsubsection{AI Usage Patterns and Tool Preferences}

Our survey revealed that all information workers who participated in our formative study were active AI tool users, with a majority using AI tools daily (81.94\%) and the remainder using them several times a week. \textit{ChatGPT} (91.67\%) was the most widely used tool, followed by \textit{Perplexity AI} (63.88\%) and \textit{Claude} (62.5\%). The prevalent usage of multiple AI tools might suggest that users maintain a suite of different assistants to accomplish various tasks, indicating no single tool fully satisfies all their individual work style preferences.

Software-wise, \textit{VS Code} (79.16\%) or other IDEs and \textit{Microsoft Suite} applications dominated participants' toolsets, illustrating the technical and document-centric nature of their productivity environments.

\begin{figure}[htbp]
  \centering
  
  \begin{subfigure}[b]{\linewidth}
    \centering
    \includegraphics[width=0.7\linewidth]{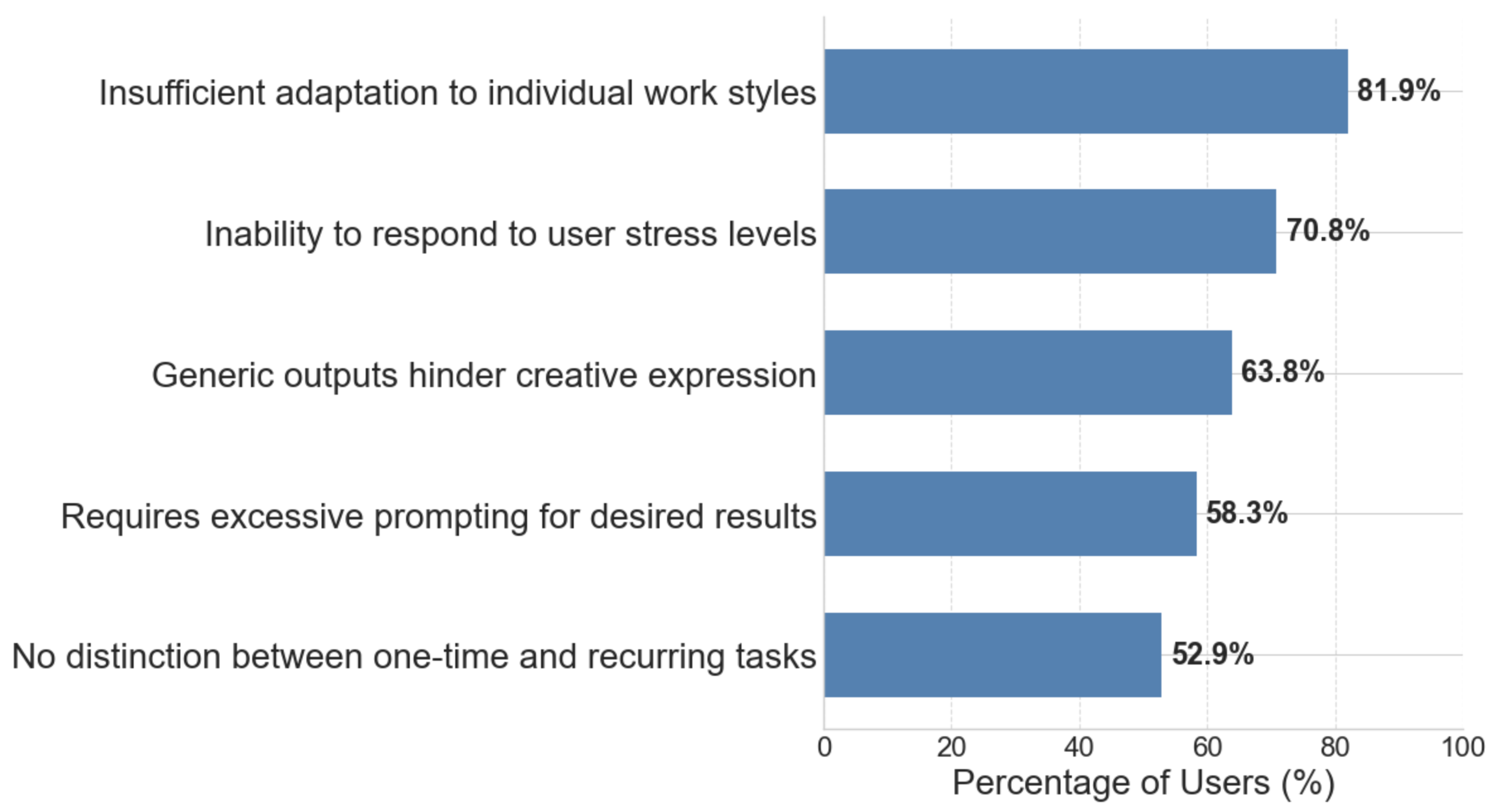}
    \caption{}
    \label{fig:part_one}
  \end{subfigure}  
  \begin{subfigure}[b]{\linewidth}
    \centering
    \includegraphics[width=0.7\linewidth]{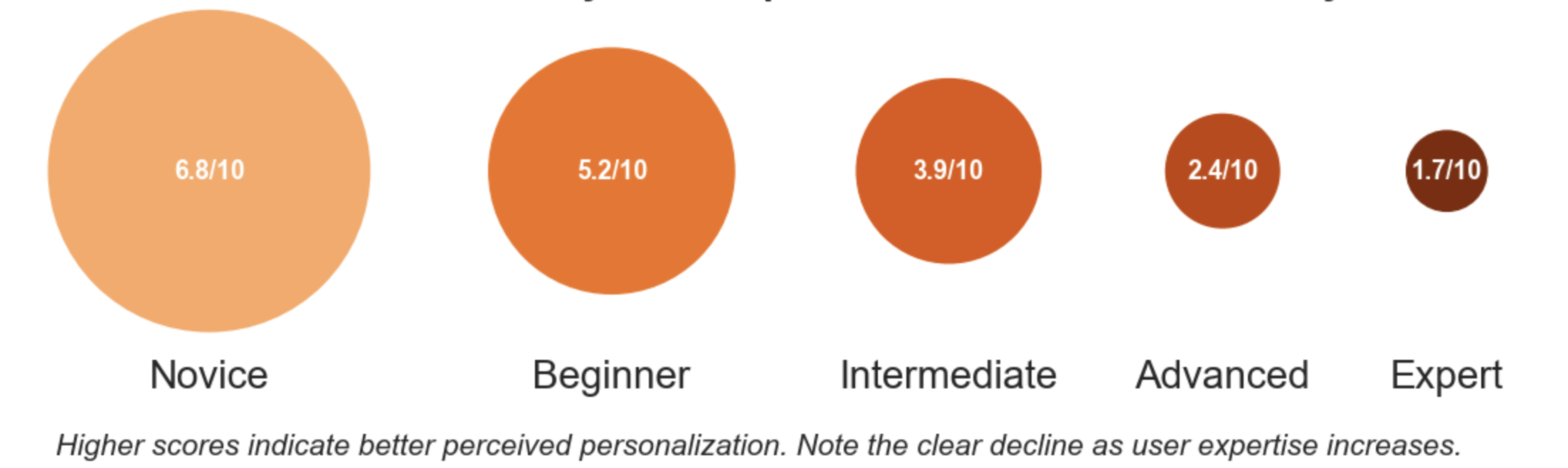}
    \caption{}
    \label{fig:part_two}
  \end{subfigure}

  \caption{(a) Perceived limitations in Personalization of Current AI Systems, and (b) Perceived Personalization by User Experience Level.}
  \Description{The image contains two visualizations. The top is a horizontal bar chart titled 'Perceived Limitations in AI System Personalization', showing common user concerns. The highest reported limitation is 'Insufficient adaptation to individual work styles' at 82.2\%, followed by 'Inability to respond to user stress levels' at 71.4\%, 'Generic outputs hinder creative expression' at 65.1\%, 'Requires excessive prompting for desired results' at 58.3\%, and 'No distinction between one-time and recurring tasks' at 52.9\%. Below this is a series of five circles showing 'Perceived Personalization by User Experience Level'. Each circle represents a user experience group from Novice to Expert, with decreasing scores: Novice (6.8/10), Beginner (5.2/10), Intermediate (3.9/10), Advanced (2.4/10), and Expert (1.7/10). The note below states that higher scores indicate better perceived personalization and highlights a clear decline as user expertise increases.}
  \label{fig:form_3}
\end{figure}

\subsubsection{Identifying Core User Needs and System Limitations}

While participants used AI for a wide range of productivity-focused activities (Figure \ref{fig:form_1}), our analysis of their challenges and the limitations of current tools revealed three critical gaps that AwareLLM is designed to address.

\begin{enumerate}
    \item \textbf{The Mind-Body Disconnect: A Need for Physical State Awareness:} A central theme was the strong, yet unassisted, connection between physical well-being and productivity. When encountering roadblocks, participants' most common coping strategies were physical interventions (Figure \ref{fig:form_2}). Taking scheduled breaks (45.8\%), hydration (38.9\%), and physical movement or stretching (33.3\%) were all more common than turning to an AI for help (26.4\%). This indicates that workers are acutely aware of their physical state (e.g., fatigue, poor posture) and actively manage it to maintain focus. However, current AI assistants are blind to this physical context. This finding provided the primary motivation to design a system capable of perceiving and responding to the user's physical state.
    \item \textbf{The Awareness Gap: Unperceived Distractions and the Demand for Proactivity:} The second major gap was the conflict between the need for focus and the reality of self-distraction. Participants reported frequently ``losing awareness of time spent on distracting activities'', making it difficult to self-regulate. This is compounded by the passive nature of current AIs. As shown in Figure \ref{fig:part_one}, a majority of users felt that AI systems require excessive prompting (58.3\%) and fail to recognize when a user is stuck. Qualitative feedback revealed a strong desire for more ``proactive, iterative assistance that can detect periods of non-productivity and automatically offer relevant help''. This demonstrated a clear user need for a system that can monitor the digital environment and proactively intervene to gently restore focus.
    \item \textbf{The Contextual Void: An Absence of Psychophysiological Awareness:} The most significant limitation identified in current AI systems was their inability to adapt to a user's internal state. A striking 70.8\% of respondents cited the ``inability to respond to user stress levels'' as a key failure of AI personalization (Figure \ref{fig:part_one}). This frustration was linked to a broader issue: a lack of adaptation to individual work styles (81.9\%) and experience levels. As Figure \ref{fig:part_two} illustrates, perceived personalization plummets as user expertise increases, from 6.8/10 for novices to a mere 1.7/10 for experts. Users expressed a clear desire for a system that understands their cognitive and emotional state—their stress, focus, and expertise—and tailors its responses accordingly. This finding established the core requirement for a system that could sense and adapt to a user's real time psychophysiological state.
\end{enumerate}

\subsection{From Formative Findings to AwareLLM’s Design}
\label{sec:design_implications}

The insights from our formative study served as a direct blueprint for the design of AwareLLM. Our findings revealed a fundamental disconnect between the static, reactive nature of current AI assistants and the dynamic, embodied reality of knowledge work. The data showed that user productivity is not merely a function of digital inputs, but is deeply intertwined with their physical state, their attentional focus, and their internal cognitive and emotional load. This pointed to a clear need for an AI assistant that could perceive and adapt to this holistic human context.

To address these interconnected needs, we established that a truly context-aware system must possess three distinct layers of awareness:

\begin{enumerate}
    \item \textbf{Physical State Awareness:} The ability to monitor the user's physical well-being, particularly ergonomic factors like posture, which our study revealed is a critical, yet unassisted, self-management strategy for 33.3\% of users.
    \item \textbf{Digital and Environmental Awareness:} The ability to understand the user's on-screen task context and perceive off-screen distractions, thereby enabling the proactive support that 58.3\% of participants requested to overcome the need for excessive prompting.
    \item \textbf{Psychophysiological Awareness:} The ability to sense the user's internal state, such as stress and cognitive load, to fill the ``contextual void'' that a striking 70.8\% of users identified as a primary failure of current AI assistants.
\end{enumerate}

These three layers of awareness form the core design philosophy of AwareLLM. To implement them, we selected a suite of sensors where each modality was chosen to directly address a specific, user-identified need. Table~\ref{tab:design_mapping} provides a clear mapping from the formative findings to the final system architecture, providing a data-driven rationale for each major component.

\begin{table*}[!htp]
    \centering
    \caption{Mapping Formative Study Findings to AwareLLM's Multimodal Sensing Architecture and Design Rationale}
    \label{tab:design_mapping}
    \begin{tabular}{p{0.35\linewidth} p{0.28\linewidth} p{0.3\linewidth}}
    \toprule
    \textbf{Identified User Need} & \textbf{Required System Layer} & \textbf{Resulting Modality} \\
    \midrule
    Users actively manage their physical state but lose track of posture during focused work. & 
    Physical State Awareness & 
    \textbf{External Webcam} for real time and continuous posture detection to provide ergonomic feedback. \\
    A majority (58.3\%) feel current AIs require excessive prompting and desire proactive, contextually relevant assistance. & 
    Digital Context Awareness & 
    \textbf{Periodic Screenshots} to enable the system to understand the user's on-screen task and provide timely, relevant suggestions. \\
    Users report “losing awareness” of time spent on off-task activities, particularly on secondary devices like phones. & 
    Environmental Awareness & 
    \textbf{Egocentric Vision} (from the eye-tracker) to detect off-task behavior in the user's immediate environment. \\
    A striking 70.8\% of users cited the AI's inability to respond to their stress levels as a critical failure of personalization. Users desire an AI that adapts to their cognitive state, avoiding overwhelming them, a need reflected by personalization scores dropping to 1.7/10 for experts. & 
    Psychophysiological Awareness & 
    \textbf{ECG Belt} to monitor Heart Rate Variability (HRV) as a robust indicator of autonomic stress and \textbf{Eye-Tracker} to infer cognitive load and attentional state via pupillometry and gaze analysis. \\
    \bottomrule
    \end{tabular}
\end{table*}

This data-driven approach ensures that every core component of the AwareLLM ecosystem is grounded in the expressed needs and challenges of knowledge workers. At the same time, our findings revealed the importance of designing with restraint. Participants valued proactive and context-aware support, yet also voiced concerns about autonomy, creativity, and the possibility of over-dependence. In response, we adopted a design philosophy centered on minimally intrusive, optional, and user-calibrated interventions. Rather than dictating actions, AwareLLM aims to gently surface awareness cues and context-appropriate assistance, ensuring that users remain in control of their workflows. This balance between proactivity and agency is essential to avoid stifling the very creativity and productivity that the system seeks to enhance.



\section{AwareLLM}
After analyzing insights from our formative study, we developed AwareLLM—a proactive, multimodal assistant designed to support user productivity in the workplace. It integrates an array of sensors with the powerful capabilities of LLMs to continuously provide personalized, real time interventions.

\subsection{Overview}
\begin{figure*}[htbp] 
    \centering
    \includegraphics[width=0.9\linewidth]{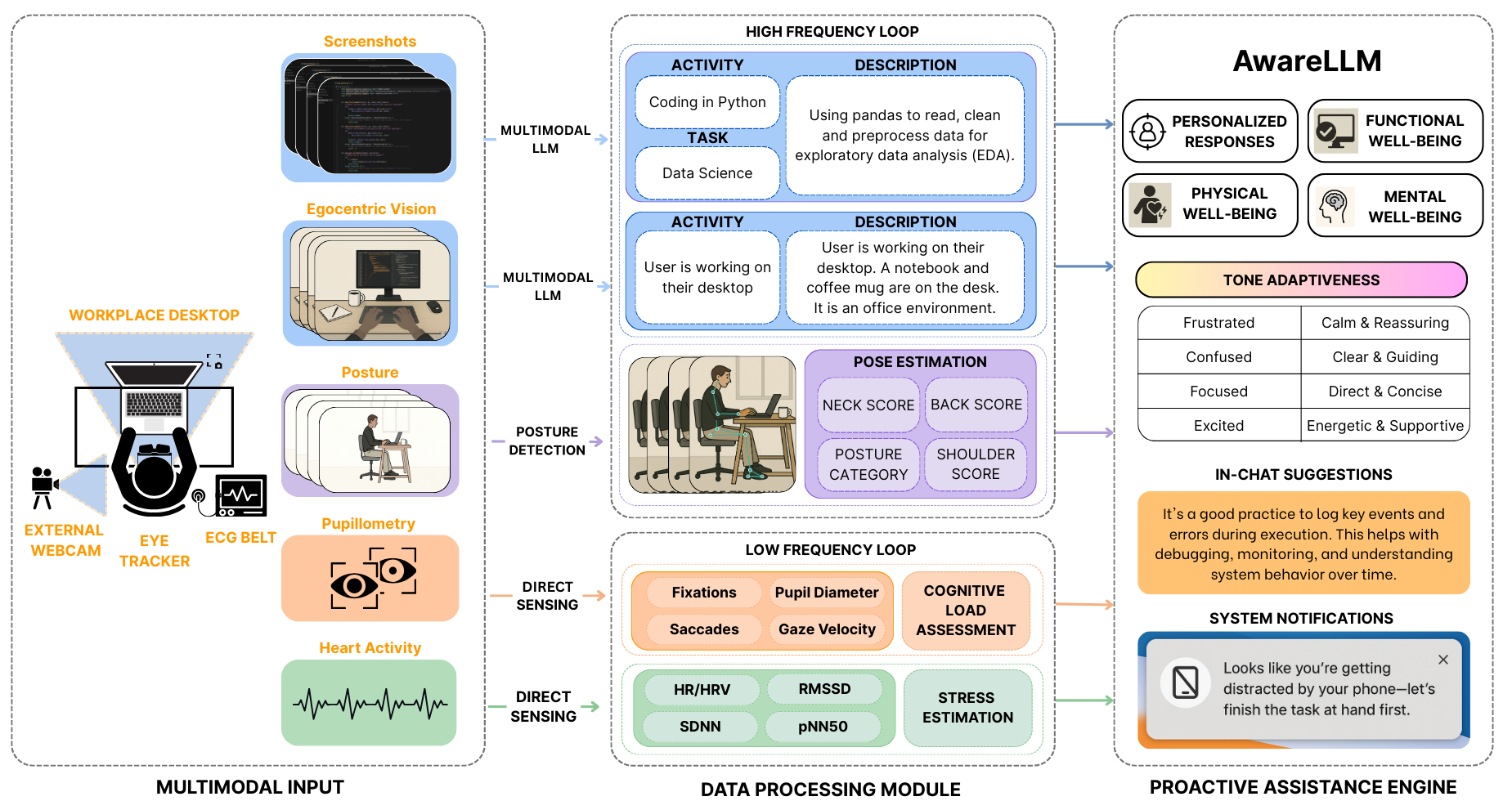} 
    \caption{AwareLLM System Architecture. The system comprises three core modules: (A) Multimodal Input, which captures visual, physiological, and contextual signals; (B) Data Processing Module with high-frequency and low-frequency loops for activity recognition, posture estimation, stress, and cognitive load assessment; and (C) Proactive Assistance Engine that delivers personalized, real time feedback to boost productivity. Assistance is provided through in-chat suggestions (task-focused) and system notifications (user-focused), with real time tone adaptation that responds to the user's emotional state. }
    \Description{A block diagram showing the architecture of AwareLLM across three main stages. On the left, under Multimodal Input, data sources include egocentric vision, screenshots, external webcam, eye tracker, and ECG belt. In the middle, the Processing Module contains a high-frequency loop (task/activity recognition and pose estimation) and a low-frequency loop (pupillometry and cardiac activity for stress and cognitive load estimation). Outputs are passed to the rightmost section labeled Proactive Assistance, where the AwareLLM agent delivers system notifications and in-chat suggestions. The assistant promotes physical, functional, and mental well-being through personalized responses.}
    \label{fig:AwareLLM}
\end{figure*}

AwareLLM, as depicted in Figure \ref{fig:AwareLLM}, is designed with the modern workplace in mind, where cognitive load, digital distraction, and physical strain can significantly impact productivity. It comprises three primary components:

\begin{enumerate}
  \item \textbf{Multimodal Input:} Integrates an external webcam, an eye tracker, an electrocardiogram (ECG) belt, and periodic system screenshots to comprehensively capture the user’s physical posture, surrounding environment, eye gaze and pupillometry, cardiac activity, and digital screen activity. 
  \item \textbf{Data Processing Module:} Extracts relevant features from these streams—skeletal pose, gaze and pupil metrics, heart rate (HR) and heart rate variability (HRV)  measures, activity classification from screenshots, and egocentric imagery via API calls to OpenAI’s lightweight model: \texttt{gpt-4o-mini}.
  \item \textbf{Proactive Assistance Engine:} Synthesizes the processed data to generate psychophysiological interventions and task-specific recommendations. It evaluates engagement levels, anticipates potential distractions, and delivers timely guidance—from posture adjustments and stress-relief cues to productivity tips such as literature review strategies or code completion suggestions.
\end{enumerate}

\paragraph{Tone-Adaptiveness} A key feature of AwareLLM’s assistance engine is tone adaptation—its ability to modulate the tone of its responses based on the user's current emotional and physiological state. By analyzing biosignals such as heart rate variability and pupil dilation, alongside behavioral cues and past chat history, AwareLLM infers the user’s affective state (e.g., frustration, fatigue, focus \cite{10.1145/2207676.2208525, saffaryazdi2022using}). It then tailors its interventions accordingly—for instance, offering a calm and empathetic reminder to take a break when stress markers are high, or adopting a more upbeat, encouraging tone when a user appears disengaged or distracted. This emotional intelligence enables AwareLLM to feel less robotic and more human-aligned in its support, fostering better receptivity and trust.

\paragraph{Chat Interface} We implemented AwareLLM as a chat-based interface to provide a familiar, user-friendly conduit for real time assistance. The interface is integrated with the desktop operating system, enabling direct system-level notifications when urgent feedback is required. By combining physical, digital, and cognitive state information, AwareLLM maintains a continuous feedback loop that enhances focus, efficiency, and well-being in professional settings.

\subsection{Multimodal Input}
AwareLLM captures and preprocesses a rich array of sensor data to form a unified representation of the user and their environment. Each modality is encoded into structured JSON objects and temporally aligned for semantic interpretation. This multimodal fusion enables AwareLLM to reason about short-term fluctuations and long-term patterns in user state, forming the foundation for real time and contextually aware support.

\begin{table}[htbp]
\caption{Multimodal Inputs and Acquisition Sources}
\label{tab:implementation_details}
\centering
\renewcommand{\arraystretch}{1.2}
\begin{tabularx}{\columnwidth}{
    >{\raggedright\arraybackslash}p{0.4\columnwidth}
    >{\raggedright\arraybackslash}X
}
\toprule
\textbf{Input Modality} & \textbf{Acquisition Source} \\
\midrule
Posture Sensing & Logitech C920 Webcam \\
Activity Classification & System Screenshots \\
Eye Tracking \& Egocentric Vision &
Pupil Labs Pupil Core Eye Tracker \\
Cardiac Activity & Maxim MAX-ECG-MONITOR Chest Belt \\
\bottomrule
\end{tabularx}
\end{table}



\paragraph{User Preferences} Customizable settings define the assistant’s tone, response style, frequency of interventions, and communication format. These parameters are embedded in the AwareLLM's system prompt, allowing it to tailor its support to individual work styles and preferences \cite{wu-etal-2025-aligning}.

\paragraph{Posture} An external webcam operating at 30 FPS captures live video, which is streamed to the posture detection module. This module employs MediaPipe’s \textit{Pose Landmarker} estimation model \cite{Lugaresi2019MediaPipeAF} to extract real time skeletal keypoints. By tracking anatomical landmarks such as the shoulders, ears, and nose across time, the system detects ergonomic deviations and posture anomalies \cite{zheng2020ergonomic}.

\paragraph{Visual Context} Egocentric worldview frames are captured periodically by the \textit{Pupil Labs Eye Tracker} \cite{Kassner:2014:POS:2638728.2641695}, providing continuous insight into the user’s physical surroundings, interactions with secondary devices, and off-screen activity \cite{HENDERSON2003498, NUNEZMARCOS2022175, POPPE2010976}. Simultaneously, high-resolution desktop screenshots capture digital context, including active applications, code editors, documents, and browser tabs \cite{imler2011screen}. Together, these visual inputs enable detailed identification of the user’s task and working environment.

\paragraph{Cardiac Activity} ECG data is collected via a chest-worn ECG Belt that communicates with the system using Bluetooth Low Energy (BLE). The belt samples at 125 Hz, delivering high temporal resolution necessary for accurate heart rate detection. The raw ECG signal is transmitted using the Lab Streaming Layer (LSL) protocol \cite{Kothe2024.02.13.580071}, ensuring real time synchronization with other sensory streams. Continuous heart rate (HR) and heart rate variability (HRV) metrics \cite{kim2018stress} derived from this stream allow AwareLLM to assess the user’s stress levels and autonomic balance over time.

\paragraph{Gaze Analysis and Pupillometry} The eye tracker, operating at 120 Hz collects raw gaze coordinates and pupil diameter measurements. This data is transmitted as a synchronized LSL stream, allowing for temporal alignment with other signals. Metrics such as pupil diameter, blink rate \cite{fogarty1989eye}, fixation duration, and saccadic activity \cite{Mayfrank1986} serve as indicators of cognitive load and attentional shifts \cite{poole2006eye}.

The selection of these specific sensors constitutes a proof-of-concept designed to directly address the multifaceted needs identified in Section~\ref{sec:design_implications}. We acknowledge that less intrusive alternatives may exist for some modalities. For instance, future iterations could explore leveraging users' existing hardware, such as smartwatch sensors for cardiac activity data or integrated laptop webcams for periodic posture estimation, trading sensor fidelity for increased ecological validity and user acceptance. However, this suite was chosen for its ability to provide rich, high-fidelity data necessary to validate our framework's core hypotheses. The significant privacy and practicality implications of such a sensing apparatus are critical limitations, which we address in detail in our discussion (Sections~\ref{sec:discussion} and \ref{sec:privacy}). All data was processed in real time and immediately deleted post-inference to protect user privacy.

\subsection{Data Processing: Bridging Sensing and Reasoning}
\label{sec:data_processing}
The Data Processing Module serves as the critical bridge between the raw, high-volume sensor inputs and the Contextual Reasoning engine. Its purpose is to transform these noisy and heterogeneous data streams into a unified, structured, and semantically-rich JSON format that the LLM can interpret. Each modality undergoes a specialized pipeline to extract key features, which are then temporally aligned and normalized to form a coherent snapshot of the user's state.

\paragraph{Posture Assessment}
To mitigate physical fatigue from prolonged desk work \cite{Amiri2024}, the system translates the webcam feed into an ergonomic score. It uses the \textit{MediaPipe Pose Landmarker} model \cite{Lugaresi2019MediaPipeAF} to detect upper-body keypoints, from which we compute three ergonomic dimensions based on established biomechanical analysis techniques \cite{Roggio2024, zheng2020ergonomic}. The final smoothed score is categorized (e.g., \textit{POOR}, \textit{IDEAL}) to provide a clear signal to the reasoning engine.

\paragraph{Visual Input Analysis}
To understand the user's task and environment, the system analyzes two visual streams. Periodic desktop \textit{screenshots} provide digital context on the user's active applications and tasks \cite{imler2011screen}. Simultaneously, \textit{Egocentric Worldview Frames} captured by the eye tracker offer insight into the user’s physical surroundings and off-screen activity, a common technique in real-world scene perception studies \cite{HENDERSON2003498, NUNEZMARCOS2022175}. These images are processed by a lightweight vision model (\texttt{gpt-4o-mini}) to extract a concise summary of the user's digital and physical context.

\paragraph{Cardiac Activity Estimation}
To infer psychological stress, the raw ECG signal is processed using a standard HRV analysis pipeline, as HRV is a robust indicator of autonomic nervous system activity and mental strain \cite{shaffer2017overview, kim2018stress}. After an initial 60-second stabilization period to minimize signal artifacts \cite{Clifford2006AdvancedMA}, a personalized baseline is established. The system then performs real time R-peak detection \cite{pan1985real} to compute key time-domain HRV metrics recommended for psychophysiological research \cite{berntson1997heart, laborde2017heart}. The final stress classification (\textit{Low}, \textit{Normal}, \textit{High}) is determined by tracking significant deviations in these metrics from the user's baseline.

\paragraph{Gaze and Pupillometry Analysis}
To estimate cognitive load, a well-established application for eye-tracking in HCI \cite{poole2006eye, Kosch2023}, the system processes raw gaze and pupil data. A dispersion-based algorithm identifies fixations \cite{salvucci2000identifying}, while saccades are detected based on gaze velocity thresholds common in eye-movement research \cite{fischer1987mechanisms}. Pupillometry data is filtered using Savitzky-Golay methods \cite{LUO20051429}, as pupil diameter is a reliable index of cognitive effort \cite{beatty1982task, vanDerWel2018Pupil}. The final cognitive load estimate is derived from a weighted synthesis of these indicators, providing a continuous score from 0--100.

\paragraph{Personalized Baselining and State Classification.}
To provide personalized, adaptive feedback, the system first needs to understand a user's unique resting state. For physiological streams like ECG and eye-tracking, the system discards the initial 60 seconds of data to allow for sensor stabilization and to minimize signal artifacts \cite{Clifford2006AdvancedMA}. It then establishes a personalized, session-specific baseline by analyzing the subsequent 60 seconds of data while the user is in a resting state. This baseline acts as a personal benchmark. High-level classifications such as ``High Stress'' or ``High Cognitive Load'' are not determined by absolute values, but by tracking the magnitude and duration of significant deviations in the processed metrics (e.g., HRV, pupil diameter) from this established baseline \cite{laborde2017heart}. This relative, baseline-driven approach is what bridges the gap between low-level raw metrics and high-level classifications, allowing the system to adapt to individual differences and day-to-day variations in a user's physiological state.

\subsection{Contextual Reasoning and Architectural Rationale}
\label{sec:contextual_reasoning}

At the core of AwareLLM lies its capacity for contextual reasoning: the ability to interpret and synthesize heterogeneous data streams from the user's environment, physiology, and digital activity to form a coherent, real time understanding of their state. This section explains why we employ an LLM rather than rigid rules, how the dual-loop reasoning pipeline operates, and how its components interact to support timely yet measured interventions.

\subsubsection{Why an LLM Instead of Rule-Based Inference?}
A naive alternative to our approach would be to use rule-based triggers (e.g., \texttt{IF posture\_score < 50 for 30s THEN notify}). Such systems are brittle: they cannot capture nuanced human behavior, often overreact to transient spikes, and fail to adapt to individual user preferences. In contrast, an LLM allows holistic reasoning across modalities and temporal windows, producing natural, tone-adaptive feedback that can incorporate user-configured preferences. This flexibility is critical for distinguishing, for instance, stress caused by a coding error (visible on-screen) versus stress caused by environmental distraction (visible in egocentric view). Hence, we employ an LLM as the central reasoning engine.

\subsubsection{The Flow of Data Through the Loops}
Once the system establishes a personalized baseline (Section~\ref{sec:data_processing}), multimodal data streams are continuously sampled and routed into two complementary reasoning loops:

\begin{itemize}
    \item \textbf{High-Frequency (HF) Loop:} Every 15 seconds, posture, screenshot, and egocentric samples are captured. Four such snapshots are aggregated into a structured JSON summary representing one minute of activity. This summary also embeds the previous minute’s trends, enabling continuity across windows. The LLM is invoked every minute with this JSON to deliver short-term context-aware interventions.
    \item \textbf{Low-Frequency (LF) Loop:} Every three minutes, the system aggregates (i) the last three HF summaries and (ii) 3-minute windows of physiological data (HRV, pupil metrics). This consolidated JSON captures both sustained psychophysiological states and evolving digital/physical context. The LLM is then invoked every three minutes to reason about longer-term patterns such as stress buildup, fatigue, or persistent distraction.
\end{itemize}

\subsubsection{Justification for the Dual-Loop Architecture}
The decision to employ a dual-loop architecture was driven by the heterogeneous temporal characteristics of the multimodal signals AwareLLM processes. Prior HCI work shows that digital interactions (e.g., on-screen behavior, posture) fluctuate rapidly, whereas physiological markers such as heart rate variability or pupil dilation evolve more slowly and require temporal smoothing to avoid false alarms \cite{shaffer2017overview, poole2006eye}. A single uniform sampling window would therefore be ill-suited to capture both short-term context and sustained states.

\begin{itemize}
    \item \textbf{Rapid Responsiveness:} The High-Frequency (HF) loop addresses fast-changing signals like posture and screen activity. Operating at a 1-minute cadence, it provides immediate situational awareness and can surface lightweight interventions that maintain flow without lag.
    \item \textbf{Deliberate Stability:} The Low-Frequency (LF) loop aggregates three HF summaries together with 3-minute windows of physiological data. This deliberate pacing mitigates the risk of overreacting to transient spikes (e.g., a single moment of stress) and supports more reliable detection of persistent cognitive or affective states \cite{laborde2017heart, Kosch2023}.
\end{itemize}

By separating immediate responsiveness from slower, confirmatory assessment, the dual-loop architecture balances \textit{sensitivity} (catching short-term issues quickly) with \textit{robustness} (avoiding transient false positives). The one-minute HF cadence aggregates four 15-second contextual observations to preserve responsiveness. The three-minute LF window is consistent with ultra-short-term HRV research using one- to three-minute segments for mental-stress detection and cognitively demanding tasks \cite{castaldo2019ultrashort,bernardes2022hrv}. We therefore treat the HF cadence as a responsiveness-oriented prototype choice and the LF duration as literature-informed, rather than claiming either is universally optimal.

\begin{figure}[htbp]
  \includegraphics[width=1.0\linewidth]{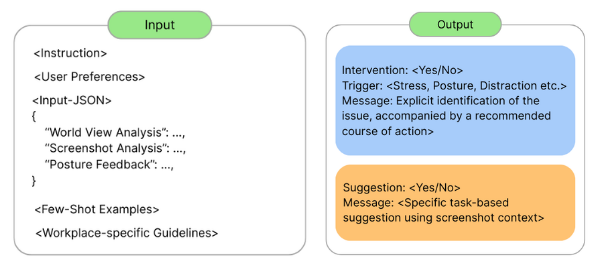}
  \caption{An illustration of the reasoning flow AwareLLM undergoes during a high-frequency loop. The model receives multimodal input including user preferences, JSON-formatted analysis from sensors, few-shot examples, and workplace-specific guidelines. It generates structured outputs in the form of interventions or task-specific suggestions based on contextual understanding.}
  \Description{A two-part diagram showing the input and output structure of the AwareLLM reasoning module. The Input section includes instructions, user preferences, a JSON object with world view analysis, screenshot analysis, and posture feedback, as well as few-shot examples and workplace-specific guidelines. The Output section includes two blocks: one for interventions, with trigger types like stress or distraction and associated recommended actions, and another for suggestions, offering task-based guidance using screenshot context.}
  \label{fig:reasoning_flow}
\end{figure}

\subsubsection{Fusion and LLM Reasoning}
The two loops are not redundant but complementary. The HF loop produces minute-level JSON summaries of posture, screen activity, and world view, each embedding trends from the previous cycle. The LF loop consolidates three such summaries with aggregated psychophysiological data to construct a broader temporal snapshot. Together, these structured inputs provide the LLM with both micro-scale context and macro-scale state trends.

The LLM is then prompted with:
\begin{enumerate}
    \item The structured JSON summary from either the HF or LF loop,
    \item User preferences (tone, frequency of intervention, notification style),
    \item Few-shot exemplars demonstrating desired reasoning strategies.
\end{enumerate}

Unlike rule-based thresholds, which have been shown to be brittle in interactive systems, the LLM can integrate cross-modal evidence and user history to infer intent and select the most contextually appropriate intervention. For example, it can distinguish stress due to task difficulty (screen content + error logs) from stress due to environmental distraction (egocentric view + gaze drift), and tailor its tone accordingly. The output is a structured intervention specification: \{\texttt{type}, \texttt{urgency}, \texttt{message}, \texttt{confidence}\}, which is then filtered through a lightweight policy layer (user preferences, do-not-disturb status, and debounce to prevent over-notification).

\subsubsection{From Architecture to User Experience}
Although the underlying reasoning involves parallel loops, multimodal fusion, and LLM inference, the complexity is deliberately hidden from the end user. From their perspective, AwareLLM appears as a familiar chat interface (Section~\ref{sec:interface}). Standard queries function as with any LLM assistant, but proactive interventions, delivered either in-chat or via system notifications, are seamlessly integrated into the workflow. This bridge from \textit{complex back-end reasoning} to a \textit{simple, chat interface} is critical: users experience AwareLLM as a supportive collaborator rather than a system that exposes its inner mechanics. In this way, the architecture supports the very goals identified in our formative study: contextual intelligence, proactivity, and minimal disruption.


\subsection{The AwareLLM Interface and Proactive Assistance Engine}
\label{sec:interface}

\subsubsection{The User's View}
\begin{figure}[htbp]
\centering
\includegraphics[width=1.0\linewidth]{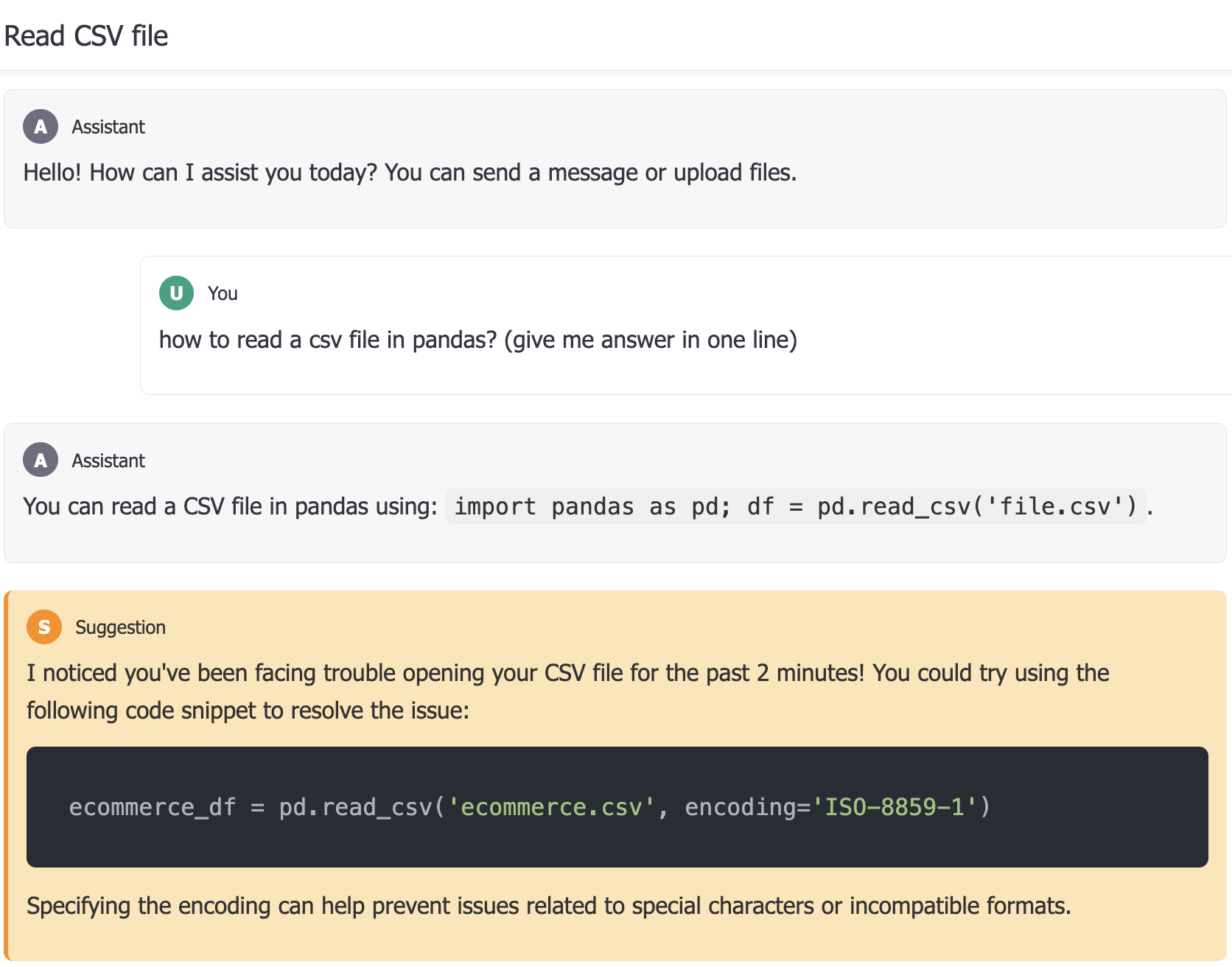}
\caption{The AwareLLM Chat Interface. While functioning as a standard chat assistant for user queries, it also proactively delivers task-specific suggestions (highlighted in orange) based on its analysis of screen activity and chat context.}
\Description{The image displays the AwareLLM chat interface. A user asks for a one-line code snippet to read a CSV. The assistant provides it. Below, in a highlighted orange box labeled ``Suggestion,'' the system proactively offers a more robust code snippet with an encoding parameter, having inferred the user might be facing issues.}
\label{fig:chat}
\end{figure}

From the user's perspective, the AwareLLM ecosystem is accessed through a familiar chat interface (Figure~\ref{fig:chat}), but it serves a dual role as both a reactive tool and a proactive partner. Users can engage with it reactively, just as they would with any standard LLM assistant, by typing direct queries. However, the system's core innovation lies in its proactive assistance engine, which initiates contact to provide timely support. A key feature that enhances both modes of interaction is the system's ability to adapt its tone based on the user's inferred state. This section details this adaptive capability, the dual-mode user interface, and the underlying intervention logic.

\subsubsection{The Proactive Assistance Engine}
The system's proactive capabilities manifest through a dual-mode delivery architecture designed to provide the right feedback through the right channel.

\paragraph{In-Chat Messages}
\sloppy
For non-urgent, task-related assistance, AwareLLM delivers suggestions directly within the chat window. These messages appear as distinctively styled, unsolicited entries (highlighted in orange) that are contextually relevant to the user's ongoing digital work. For example, if the system detects from screen activity that a user is struggling with a piece of code, it may offer a relevant code snippet, as shown in Figure~\ref{fig:chat}. This method is designed to feel like an ambient collaborator, offering subtle guidance that can be engaged with or ignored without disrupting the user's workflow.

\paragraph{System-Level Notifications}
For feedback that is more urgent or related to the user's personal well-being, AwareLLM bypasses the chat interface and uses native operating system notifications. These notifications are high-priority alerts that appear over other windows, ensuring they are immediately visible. This channel is reserved for critical user-focused feedback, such as a reminder to correct poor posture or a notification to take a break when high stress levels are detected (Figure~\ref{fig:comparison}). This separation ensures that important well-being interventions are not lost in the chat log and are given the attention they require.

\subsubsection{Intervention Logic: Supporting the Task and the User}
The dual-mode delivery system is driven by a clear logic that categorizes all proactive outputs into two distinct types. This ensures that the content of an intervention is always matched to its delivery method. Intervention selection is based on cross-modal reasoning across physiological, behavioral, and contextual evidence rather than any single sensor, informing whether, how, and through which channel AwareLLM intervenes.

\paragraph{Task-Focused Interventions (In-Chat).}
Delivered as non-urgent messages within the chat interface, these interventions are task-oriented recommendations aimed at guiding a user’s strategic approach. Generated primarily from on-screen context, chat history, and user preferences, they provide direct support for the work at hand (e.g., debugging hints, document structuring ideas), while physiological and attentional context can further inform whether such assistance should be delivered immediately or deferred.

\paragraph{User-Focused Interventions (System Notifications).}
Delivered as higher-salience system-level notifications, these interventions are designed to help the user regulate their physical or cognitive state. Rather than being triggered deterministically by individual sensor readings, they are selected when persistent multimodal evidence indicates high stress, poor posture, distraction, or sustained cognitive load.

This deliberate mapping of intervention type to delivery channel allows AwareLLM to provide a balanced real time support system. It offers subtle, task-relevant help within the user's workflow while reserving a higher-salience channel for user-focused well-being feedback. This design is informed by Interruption Management, which emphasizes minimizing the cost of poorly timed interruptions \cite{mcfarlane2002comparison}, and Attention Restoration Theory (ART), which motivates recovery-oriented support under sustained attentional demands \cite{ohly2016attention}. Together, these principles guide when and how AwareLLM intervenes while preserving user workflow and minimizing unnecessary disruption. Table~\ref{tab:interventions} provides several examples of this reasoning in practice.

\begin{table*}[t]
\caption{Theory-grounded mapping from inferred user states to AwareLLM
intervention policies and delivery.}
\label{tab:interventions}
\centering
\footnotesize
\setlength{\tabcolsep}{3.5pt}
\renewcommand{\arraystretch}{1.12}

\begin{tabularx}{\textwidth}{
    p{0.14\textwidth}
    p{0.21\textwidth}
    X
    p{0.27\textwidth}}
\toprule
\textbf{Inferred State} &
\textbf{Evidence / Persistence} &
\textbf{Theory-Grounded Rationale} &
\textbf{Delivery Policy \& Example} \\
\midrule

\textbf{Poor posture} &
Persistently low posture scores across multiple HF observations. &
Postural biofeedback can support more upright sitting during prolonged
computer work \cite{kuo2021posture}. &
\textbf{System notification.} Suppress repeated cues after a recent
intervention. \emph{``Your posture score has been low for a while, try to sit up straight.''} \\

\addlinespace

\textbf{Potential distraction} &
Repeated task-incongruent screen or environmental context during focused work session. &
Interruptions can increase stress, frustration, time pressure, and effort
\cite{mark2008interrupted}; interventions therefore require persistent,
contextually relevant evidence. &
\textbf{System notification.} Trigger only after contextual confirmation.
\emph{``Looks like you're off-task. Let's refocus on the goal!''} \\

\addlinespace

\textbf{Task impasse} &
Repeated errors, repeated actions, or limited task progress. &
Breakpoint-aware notification management can reduce disruption compared
with immediate notifications \cite{iqbal2008notifications}. &
\textbf{In-chat suggestion.} AwareLLM uses in-chat delivery as a
low-salience design choice.
\emph{``Stuck? You could try approaching this using a different library?''} \\

\addlinespace

\textbf{Elevated cognitive load} &
Sustained gaze or pupillometry deviation from the personalized baseline. &
Pupil dilation is an indirect index of cognitive effort
\cite{vanDerWel2018Pupil}; lower-workload task boundaries offer preferable
interruption opportunities \cite{bailey2008workload}. &
\textbf{Defer / in-chat.} No intervention when interruption cost
is high. \emph{``When you reach a stopping point, I can help simplify
this.''} \\

\addlinespace

\textbf{Increased physiological stress} &
Sustained HRV deviation from the personalized baseline, interpreted with
task context. &
HRV is associated with psychological stress;
attention-restoration and micro-break research motivate recovery-oriented
support \cite{kaplan1995restorative,albulescu2022microbreaks}. &
\textbf{System notification / task boundary.}
\emph{``Your stress levels seem high. Time
for a short break.''} \\

\bottomrule
\end{tabularx}
\end{table*}

\section{Experiencing AwareLLM in the Workplace}

We conducted a controlled, within-subject user study to evaluate the impact of AwareLLM on task performance and user experience in common workplace activities. The study was conducted in a lab setting with a consistent hardware and software configuration for all participants.

We recruited 20 participants (age range: 19–32; M:F = 1:1) covering diverse groups of information workers (including researchers, software developers, data analysts, and students). All participants provided informed consent via a signed form, which explicitly described the use of biosensors and data handling practices. In accordance with our privacy policy, all data points were deleted immediately after use and were never stored persistently. The experimental apparatus included:
\begin{itemize}
    \item \textbf{Sensing Apparatus:} Only two devices were body-worn: a Pupil Labs eye tracker for gaze, pupil diameter, fixations, and saccades; and a chest-worn ECG belt for heart rate and HRV metrics. Posture and body orientation were captured non-wearably using an external Logitech
    C920 webcam.
    \item \textbf{System:} The AwareLLM chat interface, with periodic system screenshots used to capture digital context.
\end{itemize}

\begin{figure*}[!ht]
    \centering
    \begin{subfigure}[b]{\textwidth}
        \centering
        \includegraphics[width=0.8\textwidth]{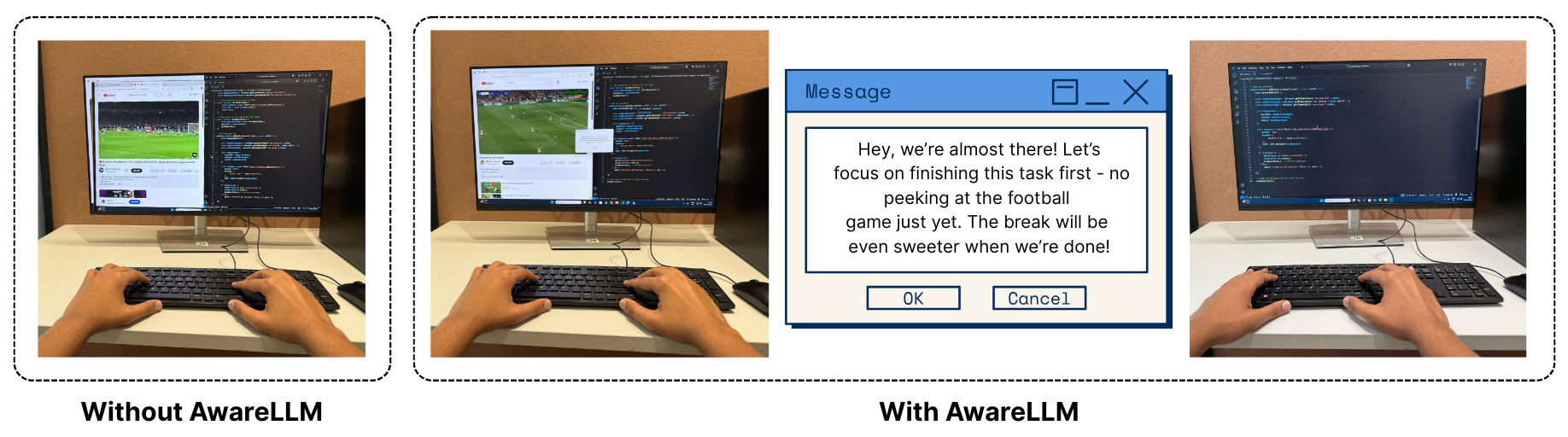}
        \caption{The user is attempting to write code while simultaneously watching a football game on YouTube.}
        \label{fig:sub1}
    \end{subfigure}
    \begin{subfigure}[b]{\textwidth}
        \centering
        \includegraphics[width=0.8\textwidth]{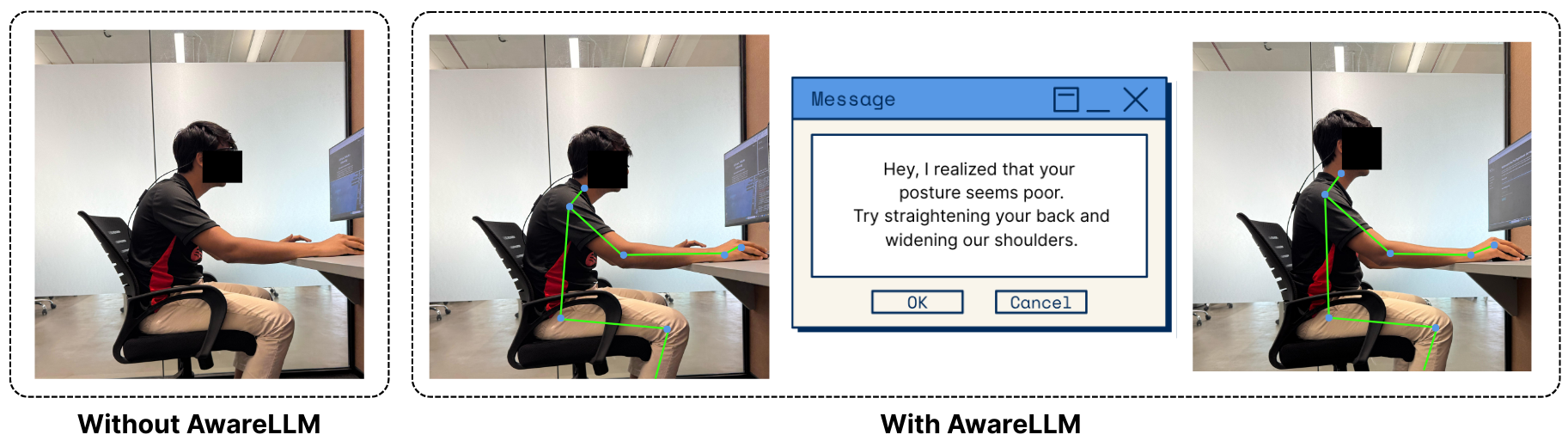}
        \caption{The user is working on their desktop but is maintaining poor posture while sitting.}
        \label{fig:sub2}
    \end{subfigure}
    \caption{Comparative Analysis of User Experience: (A) Control (Without AwareLLM) vs. (B) Treatment (With AwareLLM) Conditions in Productivity Enhancement System. In the control condition, bad posture and distractions go unnoticed and untreated, while AwareLLM notices these issues and provides appropriate real time interventions to boost user productivity.}
    \Description{This comparison image illustrates the difference between working at a computer with and without the AwareLLM system. The left panel labeled ``Without AwareLLM'' shows a user working with poor posture and becoming distracted by viewing a football game alongside their coding work. The right panel labeled ``With AwareLLM'' demonstrates the same scenarios but with intervention pop-up messages. One message advises the user to correct their posture by straightening their back and widening their shoulders, while another message encourages the user to focus on completing their primary task before watching the football game. The image effectively highlights how AwareLLM can detect and address both ergonomic issues and productivity distractions in real time.}
    \label{fig:comparison}
\end{figure*}

Each participant took part in two experimental phases—\textit{control} and \textit{treatment}—in a counter-balanced order to minimize learning effects. Ten participants experienced the control phase first, while the remaining ten began with the treatment phase. The treatment phase was equipped with the AwareLLM system, whereas the control phase used a standard LLM-based assistant (\texttt{gpt-4o-mini}) without AwareLLM functionalities. Participants were not informed about which system was active during a given session to eliminate expectation bias. Each session proceeded as follows:
\begin{enumerate}
    \item Participants completed a detailed pre-study survey, consisting of four sections:
    \begin{itemize}
        \item \textit{Participant Background:} Age, gender, role, and academic discipline.
        \item \textit{Task Experience:} Self-rated proficiency in coding, web scraping, data processing, database querying, front-end web development, and literature review.
        \item \textit{Work Style:} Preferences for task management, distraction handling, and productivity markers.
        \item \textit{AI Interaction Style:} Preferences for tone, personality traits, and proactivity of the AI agent.
    \end{itemize}
    The responses were used to personalize the behavior of the AwareLLM system.
    
    \item Participants then proceeded to interact with the assigned AI system (either during the control or treatment phase) through a dedicated chat interface. Both phases used the same underlying model, \texttt{gpt-4o-mini}, to provide responses.
    
    \item Based on our formative study results, which identified the activities where information workers most frequently seek AI assistance, we strategically selected three task categories for our experimental evaluation. The survey revealed that coding and software development (83.2\%), academic writing and research (74.6\%), and data analysis (60.1\%) were among the most common activities where participants utilized AI tools. Accordingly, participants completed the same sequence of tasks across both sessions:
    \begin{itemize}
        \item \textbf{Task 1:} Literature Review - Participants were given a research topic and asked to perform a thorough literature review, preparing a report with appropriate citations of academic sources.
        \item \textbf{Task 2:} Front-End Web Development - Participants were provided with a set of instructions to develop a web application using HTML, CSS, and JavaScript.
        \item \textbf{Task 3:} Data Science - Participants were provided with a dataset and instructed to perform exploratory data analysis (EDA) and other relevant data science tasks.
    \end{itemize}
    Each task was time-boxed to a maximum of 20 minutes. Participants were instructed to complete as much as they could within this period and were allowed to freely explore, ask the AI questions, or request assistance. 
    
    While the task categories remained consistent across all participants, the specific task instances differed between the control and treatment phases. Each pair of tasks was designed to be equivalent in difficulty and scope, ensuring a fair comparison while avoiding repetition and learning effects.


    \item \textbf{NASA-TLX.} Upon completing each task, participants filled the NASA Task Load Index (TLX) questionnaire, measuring perceived mental demand, temporal demand, effort, performance, frustration, and physical demand on an adapted 7-point scale.

    \item After the completion of all three tasks in a given phase, participants filled out a system-level survey evaluating their experience with the AI assistant. Questions were based on:
    \begin{itemize}
        \item Task manageability
        \item Perceived efficiency
        \item Ability to maintain focus
        \item Quality of output
        \item Personalization of responses
        \item Time management
        \item Overall productivity
    \end{itemize}
    These questions were rated on a 7-point Likert scale.
    
    \item After a short break, participants repeated the same set of three tasks under the other phase (control or treatment), using an entirely new task folder while maintaining the same environment and interface.
\end{enumerate}
To ensure uniformity and eliminate confounds, biosensors were actively recording during both phases. However, in the control phase, the chat interface functioned as a generic LLM-based assistant without any interventions or proactive suggestions. In contrast, the treatment phase enabled the full range of intelligent capabilities through AwareLLM (Figure \ref{fig:comparison}), including real time suggestions based on biosensor data (e.g., posture correction, cognitive load inference), proactive interventions (e.g., Pomodoro breaks \cite{10.1007/978-3-540-68255-4_18}, encouraging messages, refocusing nudges), and contextual task guidance derived from ongoing system activity.

Participants were free to interact with the AI as they wished. Their only constraint was to use the provided custom chat interface and complete the tasks to the best of their ability within the time limit. To encourage natural engagement, no guidance was given on how to prompt the AI, allowing participants to shape the interaction style themselves.

\section{Results and Findings}
To holistically evaluate AwareLLM’s effectiveness, we employed four complementary assessment methods. These were designed to capture not only task performance but also indicators of users’ mental state, perceived productivity, and objective work quality. First, we collected NASA-TLX workload ratings to measure participants’ perceived cognitive and physical demands during task completion. Second, a post-system questionnaire captured participants’ perceptions of task manageability, efficiency, and output quality. Third, an AwareLLM intervention feedback survey evaluated users’ experiences with the system’s adaptive support, including adaptation to needs, timeliness of interventions, satisfaction with assistance, stress and workload management, perceived personalization, engagement, and confidence. 

Finally, to complement these subjective assessments, we conducted an objective expert evaluation of task outputs. Industry professionals assessed the web development and data science submissions, while academic reviewers evaluated the literature review outputs, thereby providing an external benchmark of quality, structure, and completion. For all quantitative measures, statistical significance was determined using Wilcoxon signed-rank tests ($\alpha = 0.05$), as the data did not follow a normal distribution. Qualitative insights were also drawn from participant feedback collected during and after the study sessions.


\begin{figure*}[ht!]
  \centering
  \textbf{NASA-TLX Scores Comparison}\\[1ex]
  
  \begin{subfigure}[b]{0.45\linewidth}
    \centering
    \includegraphics[width=\linewidth]{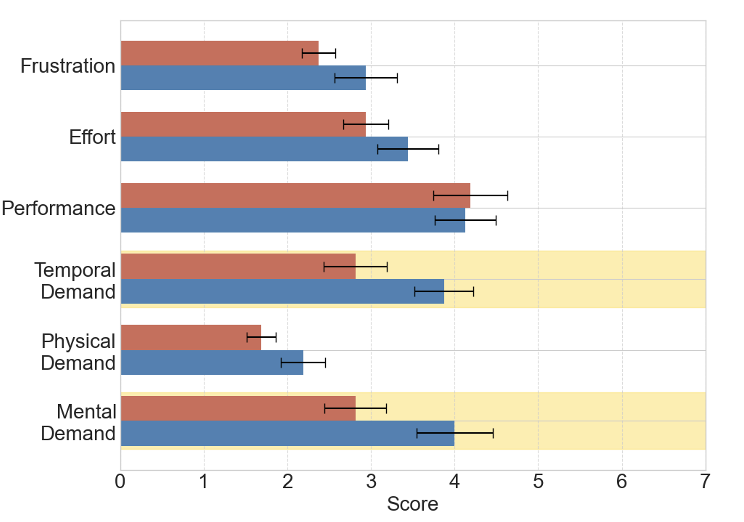}
    \caption{Literature Review}
    \label{fig:literature_review}
  \end{subfigure}
  \begin{subfigure}[b]{0.45\linewidth}
    \centering
    \includegraphics[width=\linewidth]{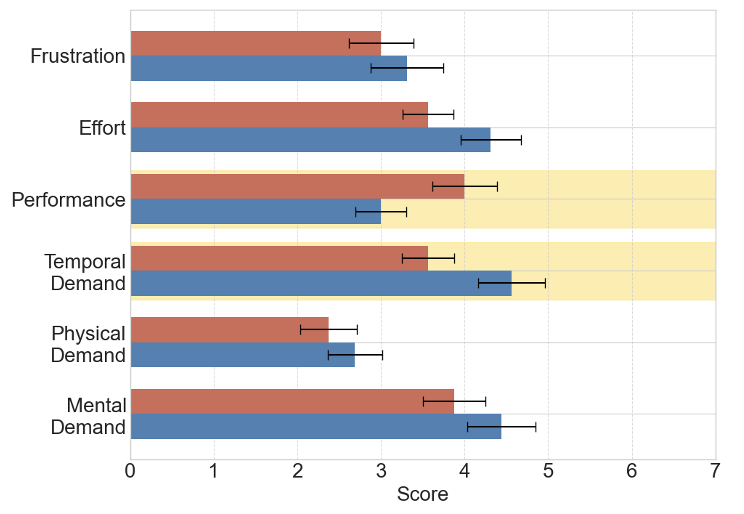}
    \caption{Front-End Web Development}
    \label{fig:frontend_web}
  \end{subfigure}

  \begin{subfigure}[b]{0.45\linewidth}
    \centering
    \includegraphics[width=\linewidth]{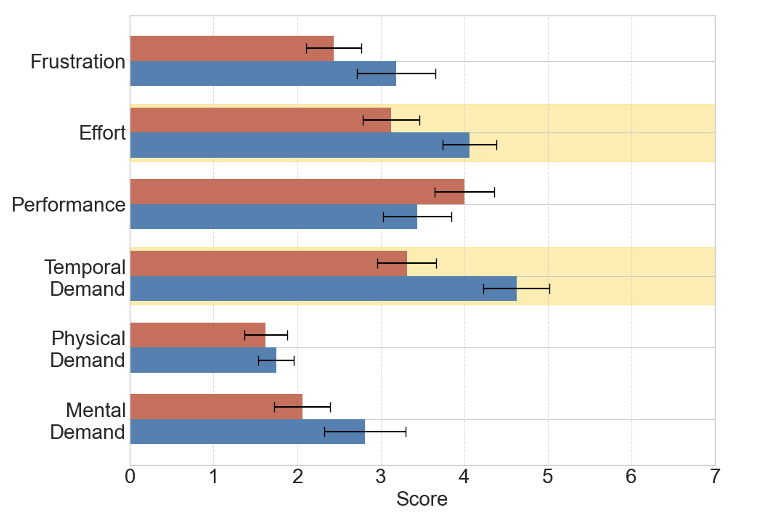}
    \caption{Data Science}
    \label{fig:data_science}
  \end{subfigure}
  \begin{subfigure}[b]{0.45\linewidth}
    \centering
    \includegraphics[width=\linewidth]{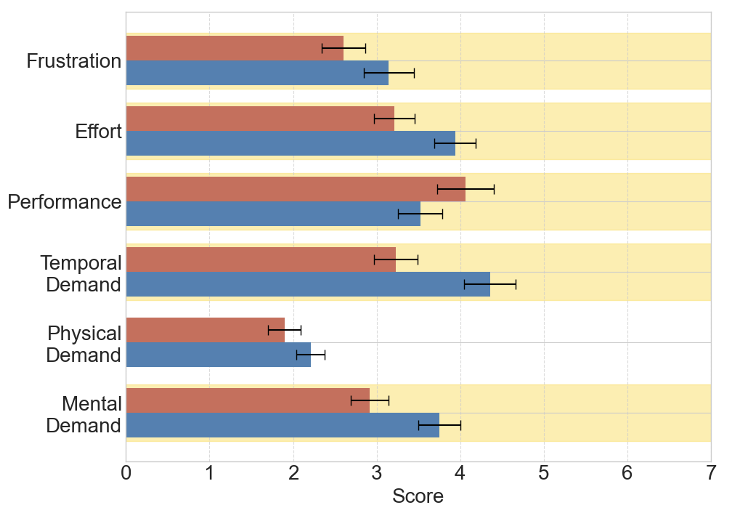}
    \caption{Overall (Mean Value Across Tasks)}
    \label{fig:overall}
  \end{subfigure}
  
  \vspace{1ex}
  \includegraphics[width=0.6\linewidth]{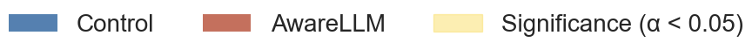}
  \Description{Legend for the objective evaluation scores figure.}
  \label{fig:legend}
  
  \caption{Participants’ ratings on NASA-TLX questions (scale: 1-low to 7-high) for the 3 tasks in the Control and AwareLLM conditions. All highlighted (in yellow) are rating differences that are statistically significant at the $\alpha = 0.05$ level via Wilcoxon signed-rank tests.}
  \label{fig:nasa_tlx_scores}
  \Description{This figure contains four grouped bar charts comparing NASA-TLX workload scores across six dimensions: Frustration, Effort, Performance, Temporal Demand, Physical Demand, and Mental Demand. The four subplots represent different tasks: (a) Literature Review, (b) Front-End Web Development, (c) Data Science, and (d) Overall Mean Value Across Tasks. Each bar chart compares scores between two conditions---Control (blue) and AwareLLM (red)---with error bars indicating variability. A shaded yellow background highlights ranges where differences are statistically significant (p < 0.05). Across all tasks, the AwareLLM condition generally shows lower Frustration, Effort, and Mental Demand compared to the Control condition, while performance scores are higher, especially in Literature Review and Data Science tasks.}
\end{figure*}

\subsection{NASA-TLX: Control vs. AwareLLM}
We conducted a detailed analysis of participants’ NASA-TLX ratings across the three task categories to evaluate the impact of AwareLLM on perceived workload. For all dimensions in the NASA-TLX ratings, lower scores indicate reduced workload, except for Performance, where higher scores reflect better self-assessed task execution. The following subsections present results disaggregated by task type, followed by a cumulative assessment aggregating responses across the entire session.
\subsubsection{Literature Review Task}
For the literature review task, as shown in Figure \ref{fig:literature_review}, AwareLLM significantly reduced Mental Demand (Control: 4.00 ± 0.46; AwareLLM: 2.81 ± 0.37; p = 0.013) and Temporal Demand (Control: 3.88 ± 0.35; AwareLLM: 2.81 ± 0.38; p = 0.021). These reductions indicate that participants experienced less cognitive demand and time constraints when using AwareLLM. Literature reviews require sustained attention and synthesis of complex information across multiple sources and can often be strenuous. Multiple participants noted that AwareLLM  suggested relevant keywords to search when they appeared stuck, highlighted important concepts across papers they were reviewing, and offered organizational frameworks that made information synthesis more manageable. 
\begin{quote}
\textit{``The system seemed to know exactly when I was struggling to connect ideas and suggested a structure that made everything clearer.''} - P9 
\end{quote}

Regarding time management, participants appreciated how they were able to work more efficiently using AwareLLM. They reported that the system would prioritize which sections to focus on, generate summaries of lengthy texts when their attention appeared to waver, and provide note-taking templates tailored to their specific literature review topic. 
\begin{quote}
\textit{``Unlike regular AI assistants that wait for you to ask for help, this system proactively stepped in when I needed it most, especially when I was feeling pressed for time.''} - P4
\end{quote}

\subsubsection{Front-End Web Development Task} In the front-end web development task (Figure \ref{fig:frontend_web}), participants reported significantly lower Temporal Demand (Control: 4.56 ± 0.40; AwareLLM: 3.56 ± 0.32; p = 0.020) and higher Performance ratings (Control: 3.00 ± 0.30; AwareLLM: 4.00 ± 0.39; p = 0.012) when using AwareLLM.
Participants highlighted how AwareLLM enhanced their coding experience through contextually aware assistance. Many noted that the system detected when they were repeatedly modifying the same code segment and offered targeted suggestions for more efficient approaches. Others appreciated how the system suggested code snippets tailored to their coding style and project requirements, reducing the need for repetitive coding.
\begin{quote}
    \textit{``It was like having a senior developer watching over my shoulder, but only stepping in when I actually needed guidance.''} - P11
\end{quote}

\subsubsection{Data Science Task} For the data science task, as indicated in Figure \ref{fig:data_science}, participants experienced significantly reduced Temporal Demand (Control: 4.62 ± 0.40; AwareLLM: 3.31 ± 0.35; p = 0.024) and Effort (Control: 4.06 ± 0.32; AwareLLM: 3.12 ± 0.34; p = 0.008) when using AwareLLM. The significant reduction in Effort was particularly emphasized in participant feedback. Some participants noted that the system detected patterns in their eye gaze that indicated confusion when examining complex datasets and automatically offered visualizations that clarified relationships within the data. 
\begin{quote}
    \textit{``It seemed to recognize when I was repeatedly scanning the same section of data and suggested better visualization to help me understand the pattern.''} \\- P18
\end{quote}




\subsubsection{Overall Performance Across All Tasks}
Figure \ref{fig:overall} presents the aggregated NASA-TLX scores across all three tasks over the session, providing a comprehensive view of AwareLLM's overall impact on workload dimensions. Using Wilcoxon signed-rank tests, we observed statistically significant improvements in five out of the six metrics:

\begin{enumerate}
    \item \textbf{Mental Demand:} Reduced by 22.1\% (Control: 3.75 ± 0.26; AwareLLM: 2.92 ± 0.22; p = 0.003)
    \item \textbf{Temporal Demand:} Reduced by 25.7\% (Control: 4.35 ± 0.31; AwareLLM: 3.23 ± 0.26; p = 0.003)
    \item \textbf{Performance:} Improved by 15.3\% (Control: 3.52 ± 0.27; AwareLLM: 4.06 ± 0.34; p = 0.029)
    \item \textbf{Effort:} Reduced by 18.5\% (Control: 3.94 ± 0.25; AwareLLM: 3.21 ± 0.24; p = 0.006)
    \item \textbf{Frustration:} Reduced by 17.5\% (Control: 3.15 ± 0.30; \\AwareLLM: 2.60 ± 0.26; p = 0.041)
\end{enumerate}

The magnitude of these improvements is particularly noteworthy. Mental Demand and Temporal Demand showed the most substantial reductions (over 22\%), suggesting that AwareLLM effectively addresses two critical bottlenecks in information work: cognitive overload and time pressure. The improvement in Performance scores, coupled with reduced Effort, indicates that participants not only felt they performed better but did so with less exertion.

While Physical Demand also showed a near-significant reduction (p = 0.091), participants did not find the tasks physically demanding as the tasks in our study were primarily cognitive in nature. 

In post-session feedback, participants generally appreciated the system’s ability to detect off-task behavior; however, some noted that AwareLLM’s focus detection could benefit from further refinement. For instance, several participants shared that when they briefly visited video streaming platforms to select background music, the system sometimes misclassified this as distraction. \textit{“The focus reminders were helpful overall”} P6 explained, \textit{“but sometimes I was just trying to find the right music to help me concentrate, and I had to dismiss the interventions.”} This highlights an opportunity to improve the system’s ability to distinguish between genuinely distracting behaviors and contextually beneficial activities that support individualized work environments.

Notably, we observed a cascading enhancement effect across the task sequence. As participants progressed through the three tasks with AwareLLM, the benefits appeared to compound. The reduced frustration and mental effort in earlier tasks left participants with greater cognitive reserves for subsequent challenges, creating a virtuous cycle of improved performance and reduced strain. This effect was particularly evident in participants' comments about feeling ``less burnt out'' and ``more energized'' to tackle later tasks compared to the control condition, where cumulative fatigue was more pronounced.

\subsection{Post-System Questionnaire: Control vs. AwareLLM} \label{sec:post_system_sec}

To comprehensively evaluate participants' experiences beyond workload assessment, we administered a post-study questionnaire using a 7-point Likert scale (1 = Not at all, 7 = Very much so). Table \ref{tab:post_system_questionnaire} presents a comparison of responses between participants using the Control system and those using AwareLLM.

The results reveal statistically significant improvements across all measured dimensions when participants used the AwareLLM system. Most notably, participants reported substantial enhancements in their ability to stay focused throughout work sessions (p = 0.009) and in the perceived quality of their work (p = 0.005). These findings align with the NASA-TLX results demonstrating reduced mental demand and improved performance across tasks.
Particularly striking was participants' perception that AwareLLM's responses felt tailored to their specific needs and workflow (p = 0.004), representing a 58\% improvement over the Control condition. This suggests that AwareLLM's context-aware approach effectively personalized assistance based on detected psychophysiological states and environmental factors, creating a more individualized experience that resonated with users.
Time management also improved significantly (p = 0.022), corroborating the NASA-TLX findings of reduced temporal demand. Participants reported feeling more capable of completing tasks efficiently and maintaining productivity throughout the session, contrasting with the cumulative fatigue observed in the Control condition.
Overall productivity showed a substantial 33\% improvement (p = 0.007), indicating that participants not only felt subjectively better about their experience but also perceived tangible benefits to their productivity when using AwareLLM. 

\begin{table*}[!htp]
  \caption{
Post-System Questionnaire Results: Comparison of responses from participants in the Control condition (without AwareLLM) and the treatment condition (with AwareLLM). Each statement was rated on a 7-point Likert scale ranging from 1 (\textit{Not at all}) to 7 (\textit{Very much so}). Reported values are means with standard deviations. Statistical significance was evaluated using the Wilcoxon signed-rank test; bolded p-values indicate differences between conditions are significant at the $\alpha = 0.05$ level.
}
\label{tab:post_system_questionnaire}
  \begin{tabular}{p{0.43\linewidth}ccc}
    \toprule
    \textbf{Claim} & \textbf{Control Mean $\pm$ SD} & \textbf{AwareLLM Mean $\pm$ SD} & \textbf{p-value}\\
    \midrule
    I found my tasks more manageable with the AI agent's assistance. & 3.44 $\pm$ 0.51 & \textbf{4.44 $\pm$ 0.48} & \textbf{0.045} \\
    The AI agent helped me work more efficiently. & 3.06 $\pm$ 0.45 & \textbf{4.00 $\pm$ 0.49} & \textbf{0.038} \\
    I was able to stay focused throughout my work session with the AI agent. & 2.75 $\pm$ 0.36 & \textbf{4.25 $\pm$ 0.51} & \textbf{0.009} \\
    The quality of my work improved with the AI agent's assistance. & 3.62 $\pm$ 0.42 & \textbf{4.81 $\pm$ 0.33} & \textbf{0.005} \\
    The AI agent's responses felt tailored to my specific needs and workflow. & 3.00 $\pm$ 0.27 & \textbf{4.75 $\pm$ 0.32} & \textbf{0.004} \\
    I managed my time effectively while completing the tasks. & 3.44 $\pm$ 0.44 & \textbf{4.75 $\pm$ 0.27} & \textbf{0.022} \\
    Overall, I feel that my productivity improved after using the AI agent. & 3.75 $\pm$ 0.31 & \textbf{5.00 $\pm$ 0.27} & \textbf{0.007} \\
    \bottomrule
  \end{tabular}
\end{table*}

The questionnaire results reveal a fundamental shift in how users perceive AI assistance. When using AwareLLM, participants did not just complete tasks, they experienced a qualitatively different relationship with their work. The consistent improvement across all metrics suggests that participants recognized and valued the system's ability to understand their needs contextually rather than just responding to explicit requests.

What distinguishes AwareLLM from conventional AI systems is its ability to bridge the gap between technological capability and human experience. Rather than treating productivity as merely a measure of output, AwareLLM acknowledges the human factors that underpin effective work: focus, perceived quality, and satisfaction. 

\subsection{AwareLLM Intervention Feedback}
To gain deeper insights into users' experiences with AwareLLM's proactive interventions, we collected feedback on seven key dimensions using a 7-point Likert scale as employed earlier in \ref{sec:post_system_sec}. Table \ref{tab:intervention_feedback} presents a summary of these responses.


\begin{table*}[!htp]
  \caption{Additional Feedback Responses from the AwareLLM Condition: The table highlights the mean and standard deviation (Mean/SD) of responses exclusively from the treatment condition, evaluating their perceptions of personalized interventions, system responsiveness, ease of interaction, the impact of interventions, and satisfaction with the assistance provided by AwareLLM}
  \label{tab:intervention_feedback}
  \centering
  \begin{tabular}{p{0.68\linewidth}cc}
    \toprule
    \textbf{Feedback Dimension} & \textbf{Mean} & \textbf{Standard Deviation} \\
    \midrule
    The AI system effectively adapted its recommendations to my changing needs. & 5.39 & 0.99 \\
    The AI provided its interventions at appropriate moments in my workflow. & 5.50 & 0.67 \\
    I am satisfied with the assistance provided by the personalized AI system. & 5.72 & 0.92 \\
    The AI's interventions helped me reduce stress and workload effectively. & 5.61 & 1.10 \\
    The AI's interventions felt uniquely tailored to my personal work style and goals. & 5.47 & 1.33 \\
    The AI's suggestions increased my engagement or motivation while working. & 6.16 & 0.84 \\
    The AI's interventions boosted my confidence in handling complex tasks. & 5.53 & 0.94 \\
    \bottomrule
  \end{tabular}
\end{table*}

Participants rated AwareLLM highly across all intervention dimensions, with particularly strong scores for Engagement and Motivation (M=6.16, SD=0.84). This suggests that the system's psychophysiologically-driven interventions not only provided technical assistance but also enhanced users' emotional and motivational states while performing complex knowledge work. As P15 noted, \textit{``It was so cool to see that the system seemed to know exactly when I needed encouragement versus when I needed specific technical guidance. It felt like it was reading my confidence levels and responding accordingly.''} The relatively low standard deviation for Timeliness of Interventions (SD=0.67) indicates consistent agreement among participants that AwareLLM delivered support at appropriate moments—neither too early nor too late in their problem-solving process. The strong score for Personalization Perception (M=5.47) demonstrates that participants recognized and valued the system's adaptation to their individual working styles, cognitive states, and physiological responses. These findings align with our NASA-TLX results and post-system questionnaire data, further supporting the effectiveness of AwareLLM's context-aware, multimodal approach to human-AI collaboration.

\subsection{Expert Evaluation: Control vs. AwareLLM}

\begin{figure*}[ht!]
  \centering
  \textbf{Expert Evaluation}\\[1ex]

  \begin{subfigure}[b]{0.45\linewidth}
    \centering
    \includegraphics[width=\linewidth]{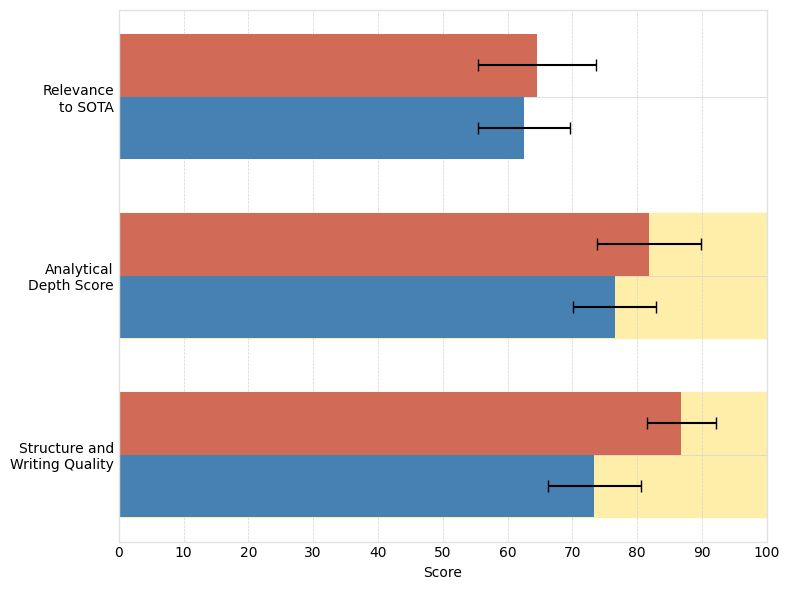}
    \caption{Literature Review}
    \label{fig:literature_review_objective}
  \end{subfigure}
  \hfill
  \begin{subfigure}[b]{0.45\linewidth}
    \centering
    \includegraphics[width=\linewidth]{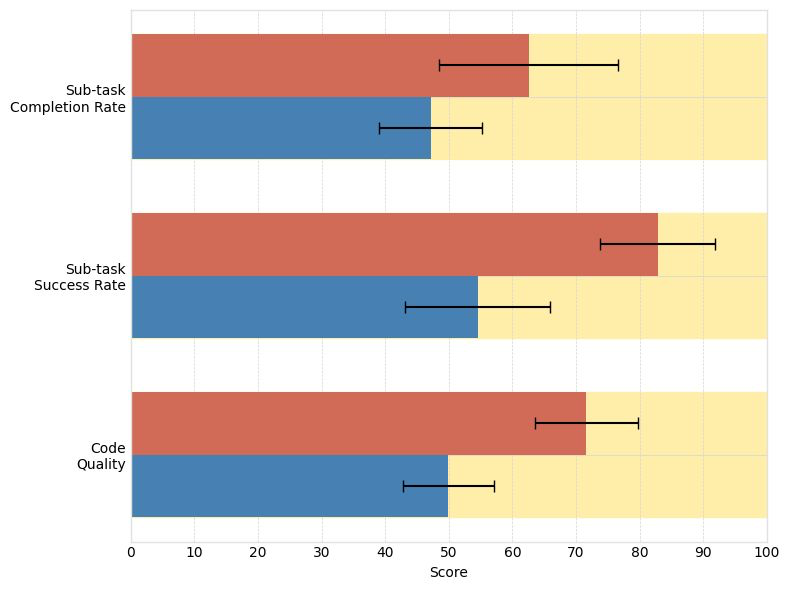}
    \caption{Front-End Web Development}
    \label{fig:frontend_web_objective}
  \end{subfigure}

  \begin{subfigure}[b]{0.45\linewidth}
    \centering
    \includegraphics[width=\linewidth]{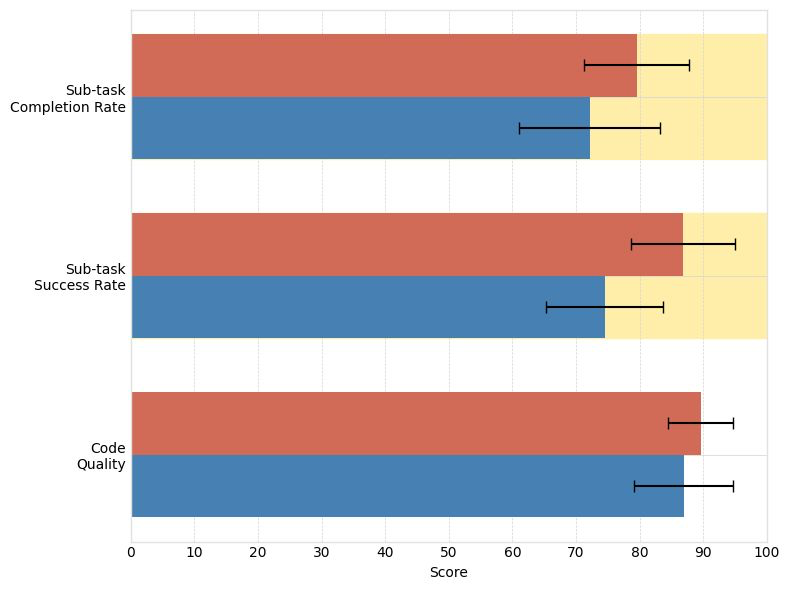}
    \caption{Data Science}
    \label{fig:data_science_objective}
  \end{subfigure}

  \vspace{1ex}
  \includegraphics[width=0.6\linewidth]{images/legend_unite.png}
  \Description{Legend for the objective evaluation scores figure.}
  \label{fig:legend2}

  \caption{Objective expert evaluation scores (scale: 0–100) for the three tasks across Control and AwareLLM conditions. Highlighted regions indicate statistically significant differences at $\alpha = 0.05$ based on Wilcoxon signed-rank tests.}
  \label{fig:objective_scores_centered_legend}
\end{figure*}
To complement the subjective and psychophysiological analyses, we conducted an objective evaluation of participants’ task outputs to provide an external benchmark of work quality, completion, and effectiveness. Unlike self-reported measures, these assessments were independently performed by a panel of two to three domain experts. Industry professionals evaluated the data science and web development submissions, while academic reviewers assessed the literature review responses. Evaluators were blinded to the experimental condition, and all submissions were anonymized. 

Each deliverable was scored across three objective performance dimensions: completion effectiveness, task-specific success, and overall quality of output, on a 0–100 scale. Higher scores represent superior task performance.
\subsubsection{Literature Review Task}
Academic reviewers evaluated the literature review submissions based on the relevance of references to the state of the art, structural coherence, and depth of synthesis and argumentation (Figure \ref{fig:literature_review_objective}).
Since the chatbots operated without live internet access, participants had to manually locate and assess research papers, leading to no significant difference between the two conditions in the relevance of references to the state of the art (Control: 62.49 ± 7.07; AwareLLM: 64.54 ± 9.12; p = 0.545).

However, AwareLLM users achieved significantly higher scores in writing structure and overall quality (Control: 73.41 ± 7.20; AwareLLM: 86.79 ± 5.33; p = 0.0002) and analytical depth (Control: 76.51 ± 6.41; AwareLLM: 81.83 ± 8.05; p = 0.048).
Reviewers noted that AwareLLM-assisted participants exhibited smoother logical flow, clearer transitions, and more balanced integration of prior works.
\begin{quote}
\textit{``While both groups selected similar papers, the AwareLLM users articulated their reasoning more clearly and presented arguments with greater cohesion.''} - AR2
\end{quote}
AwareLLM enabled participants to express ideas more naturally while proactively enhancing structural clarity and analytical depth.
\subsubsection{Front-End Web Development Task}
Industry experts found that participants using AwareLLM developed web interfaces that were cleaner, more consistent, and functionally complete than those in the control condition.
Participants completed a greater number of sub-tasks within the allotted time, and the outputs of each completed task were more robust and stable, demonstrating clear gains in both quantity and quality of work.
Projects completed with AwareLLM showed significant improvements across all three metrics: Sub-Task Completion Rate (Control: 47.16 ± 8.07; AwareLLM: 62.54 ± 14.12; p = 0.001), Sub-Task Success Rate (Control: 54.51 ± 11.38; AwareLLM: 82.83 ± 9.05; p = 0.0002), and Code Quality of Output (Control: 49.92 ± 7.11; AwareLLM: 71.61 ± 8.13; p = 0.0002) (Figure \ref{fig:frontend_web_objective}).

Experts noted that AwareLLM’s proactive debugging cues, context-sensitive prompts, and layout suggestions enabled participants to resolve issues faster and maintain coherent code architecture.
\subsubsection{Data Science Task}
As shown in Figure \ref{fig:data_science_objective}, expert evaluations revealed that code quality was strong across both conditions, as participants in the control and AwareLLM conditions used the same underlying chatbot model (Control: 86.92 ± 7.74; AwareLLM: 89.61 ± 5.13; p = 0.294). However, when it came to insight generation and analytical robustness, including exploratory data analysis, visualization, and statistical testing, AwareLLM’s proactive suggestions and contextual tips provided a clear advantage. Participants using AwareLLM achieved higher Sub-Task Completion Rates (Control: 72.16 ± 11.07; AwareLLM: 79.54 ± 8.21; p = 0.039) and Sub-Task Success Rates (Control: 74.51 ± 9.20; AwareLLM: 86.83 ± 8.14; p = 0.0003), reflecting more thorough and interpretable analyses.
\begin{quote}
\textit{``AwareLLM users approached analysis more systematically, leading to deeper and more comprehensive insights.''} - DR3 
\end{quote}
As shown in the previous sections, AwareLLM’s ability to reduce frustration, effort, and mental demand not only led to higher satisfaction ratings among participants, but also translated into objectively higher task quality, completeness, and analytical rigor across all evaluated domains.


\section{Discussion}
\label{sec:discussion}

Building upon the promising results observed across quantitative metrics and user feedback, this section delves into the broader implications of AwareLLM’s performance, design choices, and potential for real-world deployment. While the system demonstrated significant benefits in enhancing productivity, reducing mental workload, and fostering user engagement, these outcomes raise important considerations regarding the scalability, generalizability, and ethical use of multimodal, context-aware AI systems. We reflect on these aspects, examining both the strengths and the critical challenges that emerged throughout the study.

\subsection{Implications for Proactive and Embodied AI}
Our findings contribute to the growing body of work on context-aware systems in HCI \cite{Hong20098509, Dey2001}. While recent systems have effectively leveraged environmental and digital context to enhance user interaction \cite{Lee2024GazePointAR, Lindlbauer2019}, AwareLLM's primary contribution is the deep integration of physiological and embodied signals to create a more holistic model of the user \cite{Schmidt2016, Moge2022}. By fusing data on a user's physical posture, cognitive load, and autonomic stress with their on-screen activity, our work takes a critical step toward filling the ``Contextual Void,'' ``Awareness Gap,'' and ``Mind-Body Disconnect'' that our formative study (Section~\ref{sec:design_implications}) identified as key limitations of current AI tools.

\paragraph{Navigating the Proactivity Dilemma.}
A central challenge in this domain is navigating the trade-off between helpful assistance and unwanted disruption \cite{Meurisch2020}. Proactive interventions, if not carefully designed, risk undermining user agency or interrupting valuable learning processes that arise from productive struggle. Our dual-mode intervention system is a first step toward mitigating this, but future work must more deeply explore the long-term impact on skill acquisition. An important research direction is to design interventions that act as adaptive scaffolds—providing more explicit guidance for novices while fading to subtle nudges for experts, thereby supporting rather than replacing the learning process. Investigating how AwareLLM's proactivity can be tuned to prevent over-reliance and foster user competence over time is a critical next step for this line of research. Here, AwareLLM's architecture offers a key insight. Its proactivity is managed through a dual-mode delivery system (Section~\ref{sec:interface}) that distinguishes between different types of support. This approach is a deliberate attempt to manage the proactivity dilemma; by delivering task-focused suggestions as subtle in-chat messages, we preserve user agency in their primary workflow, while reserving high-priority system notifications for critical well-being nudges—akin to a physical indicator like the FlowLight \cite{Zuger2017}. This separation demonstrates how proactive systems can be designed to be powerful yet respectful of the user's focus and autonomy.

\paragraph{A New Paradigm for Human-AI Collaboration.}
Ultimately, this research pushes the boundaries of human-AI collaboration beyond simple prompt-response interactions or conditional delegation \cite{Lai2022}. By endowing the AI with an awareness of the user's embodied state, AwareLLM fosters a more symbiotic partnership. The AI is no longer just a tool to be commanded but a partner that can anticipate needs and offer support that aligns with the user's cognitive and physiological capacity. This model of a physiologically-aware assistant has significant implications for future applications in domains like personalized learning \cite{AttentivU2018}, developer tooling, and adaptive virtual environments \cite{Chiossi2023}, creating a blueprint for AI systems that support the whole person, not just the task at hand.

\subsection{Limitations and Future Work}
\paragraph{Privacy and Ethical Considerations.}
Given the use of continuous visual, physiological, and behavioral monitoring, privacy remains a central concern and a key limitation of our solution. The integration of external webcams, desktop screen capture, and biosensors introduces potential risks related to surveillance, data misuse, and participant discomfort. Although our study was conducted entirely in a controlled lab environment with informed consent, deploying such systems in real-world scenarios would require rigorous safeguards --- such as on-device processing, anonymization protocols, and transparency in data usage.

\paragraph{Interpretation Reliability and Contextual Ambiguity} A notable limitation of our framework is the inherent ``black box'' nature of LLM-based reasoning, a well-documented challenge in human-centered explainable AI (HCXAI) \cite{ehsan2024human}. While prompt engineering guides the Vision-Language Model's interpretation of egocentric frames, occasional misclassifications can occur. Furthermore, during focused work sessions, environmental actions such as prolonged smartphone usage may be inferred as potential distractions. To minimize privacy and security risks, the evaluated prototype intentionally did not analyze content displayed on participants' phones. Consequently, task-relevant phone use (e.g., testing a mobile application or reading a work-related message) may occasionally be misclassified. Future iterations could explore explicit opt-in, privacy-preserving context signals to distinguish off-task behavior from productive cross-device workflows.

\paragraph{Modularity and Sensor Generalizability.}
A key limitation of this study is the evaluation of AwareLLM as a monolithic system. A critical next step is to investigate the independent and combined contributions of each modality. Future work should conduct ablation studies to answer questions such as: How effective is the system with only posture and screen-capture data versus the full sensor suite? This would not only clarify the contribution of each signal but also allow organizations to adopt subsets of the framework that align with their specific privacy policies and available hardware. This research path is essential for transitioning AwareLLM from a lab-based proof-of-concept to a flexible, deployable framework.

\paragraph{Limited Temporal Scope.}
Our study is cross-sectional in nature: each participant engaged with AwareLLM for a single lab session. While this design allows for initial validation and interaction modeling, it does not capture behavioral trends, adaptation effects, or sustained cognitive responses over multiple work sessions. A longitudinal study would be necessary to understand how users’ productivity, trust, and interaction patterns evolve over time, and whether AwareLLM can adapt its suggestions meaningfully over days or weeks.

\section{Privacy Concerns}
\label{sec:privacy}

This research was conducted with a strong commitment to ethical standards, ensuring the protection of participants' rights and well-being. All volunteers provided informed consent before participating and retained the right to withdraw at any point without consequence. To safeguard privacy, all data were anonymized using unique identifiers, securely stored, and access was restricted to authorized personnel only. All raw and processed data collected in real time from biosensory devices were deleted immediately after use, in accordance with data minimization principles. Furthermore, the study was reviewed and approved by the hosting organization, which granted explicit permission to carry out the experiment under these ethical guidelines.

\section{Conclusion}

In this paper, we introduced AwareLLM, a unified system that leverages biosignals, contextual cues, and LLMs to support users through real time, adaptive interventions. Grounded in insights from a formative study and experimental evaluation through a user study, our design underscores the value of integrating physiological and behavioral signals to gauge cognitive state and task engagement. We presented a comprehensive framework that operationalizes attention-aware assistance through multimodal inputs and demonstrated the effectiveness of our approach through controlled experiments. Our findings show that AwareLLM significantly reduces user workload, enhances focus, and improves task performance. By showcasing the synergistic potential of human-AI collaboration, this work marks a crucial step toward intelligent systems that understand and respond to human needs with empathy and context. We envision this direction fostering a new generation of adaptive interfaces that are not only reactive but genuinely collaborative --- empowering users across diverse environments and cognitive states.


\section*{GenAI Usage Disclosure}
In compliance with the ACM Policy on Authorship, we disclose the use of generative AI tools during the research and manuscript preparation process. These tools were employed for limited purposes, including creating select graphic illustrations, refining grammar and phrasing, and generating boilerplate code templates used in participant assessments for the web development and data science tasks.

\section*{Ethical Considerations}
This study involving human participants was reviewed and approved by the Institutional Ethics Committee in accordance with the applicable ethics review procedures. All participants provided informed consent prior to their involvement, and all procedures adhered to relevant institutional and national research ethics guidelines. No personally identifiable information (PII) was collected during the experiments, and all data were immediately deleted upon use.




\bibliographystyle{ACM-Reference-Format}
\bibliography{mybibliography}
\appendix
\input{supplementary_use}

\end{document}

%% file: supplementary_use.tex


\clearpage

\twocolumn[
\begin{center}
    {\Huge\bfseries Appendix}
    \vspace{1em}
\end{center}
]




\section{Experimental Tasks}
This section outlines the specific tasks assigned to participants during the experiment, categorized by type (Literature Review, Front-end Web Development, Data Science) and variation (Control, Treatment).

\subsection{Literature Review}

\subsubsection{Control}

\paragraph{\textbf{Title:}} Lake Size Estimation Techniques
\paragraph{\textbf{Description:}} Comprehensive review of satellite imagery-based lake size estimation methods.
\paragraph{\textbf{Task Details:}} Prepare a concise academic literature review (approx. 500 words). This involves identifying and describing 3-5 cutting-edge methods for lake size tracking using satellite imagery, comparing traditional remote sensing algorithms with machine learning approaches, analyzing key performance indicators and technological limitations, and highlighting the most relevant datasets used in this field. The review must include a minimum of 5 peer-reviewed sources published after 2018.
\paragraph{\textbf{Deliverable(s):}} Structured academic report with clear section headings and citations.

\subsubsection{Treatment}
\paragraph{\textbf{Title:}} Aerial Tree Detection and Classification
\paragraph{\textbf{Description:}} Comprehensive analysis of tree inventorization methods from aerial imagery.
\paragraph{\textbf{Task Details:}} Develop a targeted literature review (approx. 500 words). The focus should be on examining 3-5 state-of-the-art methods for tree detection using aerial imagery, contrasting traditional GIS techniques with modern data-driven approaches, evaluating performance metrics and technological constraints, and highlighting the most relevant datasets for tree detection studies. The review must cite at least 5 recent (post-2018) scientific publications.
\paragraph{\textbf{Deliverable(s):}} Professionally formatted academic report with critical analysis.

\subsection{Front-end Web Development}

\subsubsection{Control}
\paragraph{\textbf{Title:}} Trivia Quiz and Weather App
\paragraph{\textbf{Description:}} Develop two interactive and fully responsive front-end applications: a Trivia Quiz Game and a Weather App.
\paragraph{\textbf{Task Details:}} Build two front-end web applications:
\begin{enumerate}
    \item \textbf{Trivia Quiz Web App:}
    \begin{itemize}
        \item Allow a user to input multiple-choice questions with a single correct answer.
        \item Enable another user to play the quiz, selecting answers while their score updates in real time.
        \item Implement scoring: +10 points for correct answers, -5 points for wrong answers.
        \item Play celebratory sounds for correct answers and error sounds for wrong answers.
        \item Show only the final score on the entire page after the last question.
        \item Ensure a clean and user-friendly UI.
    \end{itemize}
    \textit{Bonus features:}
    \begin{itemize}
        \item Add animations for correct and wrong answers.
        \item Make it responsive and visually appealing.
        \item Store quiz questions in \texttt{localStorage} for persistence.
    \end{itemize}
    
    \item \textbf{Responsive Weather App:}
    \begin{itemize}
        \item Allow users to enter a city name and display the current day's weather details.
        \item Include temperature (with Celsius and Fahrenheit toggle), humidity, wind speed, and weather conditions.
        \item Display relevant weather icons.
        \item Implement a Bootstrap-based fully responsive layout.
        \item Include a navigation bar (Home, About, Settings) that collapses into a burger menu on small screens.
        \item Provide a loading spinner during data fetching.
    \end{itemize}
    \textit{Bonus features:}
    \begin{itemize}
        \item Implement dark mode using Bootstrap's theme.
        \item Save the last searched city in \texttt{localStorage}.
        \item Change background images dynamically based on weather conditions.
    \end{itemize}
\end{enumerate}

\paragraph{\textbf{Deliverable(s):}} Fully functional front-end websites with clean, intuitive user interfaces.

\subsubsection{Treatment}
\paragraph{\textbf{Title:}} Word Guessing Game and Movie Finder App
\paragraph{\textbf{Description:}} Develop two interactive and fully responsive front-end applications: a Word Guessing Game and a Movie Finder App.
\paragraph{\textbf{Task Details:}} Build two front-end web applications:
\begin{enumerate}
    \item \textbf{Word Guessing Game:}
    \begin{itemize}
        \item Display a scrambled word that the user must unscramble by typing the correct word into an input box.
        \item Include a 'Submit' button for user responses.
        \item Implement scoring system: +10 points for correct answers, -5 points for wrong answers.
        \item Play appropriate celebratory sounds for correct answers and error sounds for wrong answers.
        \item Display real-time score updates at the top of the screen.
    \end{itemize}
    \textit{Bonus features:}
    \begin{itemize}
        \item Add difficulty levels (Easy, Medium, Hard) affecting word complexity.
        \item Allow users to enter custom words for others to guess.
        \item Implement animations for correct/incorrect answers.
    \end{itemize}
    
    \item \textbf{Movie Finder App:}
    \begin{itemize}
        \item Find and use an API (like the Open Movie Database) to fetch movie details.
        \item Allow users to search for a movie by name and display details including title, year, genre, director, plot summary, IMDB rating, and poster.
        \item Show a `Not Found' message if no movie matches the search.
        \item Implement a Bootstrap-based fully responsive layout.
        \item Include a navigation bar (Home, Search, Favorites) that collapses on mobile screens.
        \item Allow users to favorite movies, storing them in \texttt{localStorage}.
    \end{itemize}
    \textit{Bonus features:}
    \begin{itemize}
        \item Display movie recommendations based on genre.
        \item Add a loading spinner while fetching data.
        \item Implement Bootstrap's dark mode.
    \end{itemize}
\end{enumerate}

\paragraph{\textbf{Deliverable(s):}} Fully functional front-end websites with clean, intuitive user interfaces.

\subsection{Data Science}

\subsubsection{Control}
\paragraph{\textbf{Title:}} Customer Churn Analysis and Predictive Insights
\paragraph{\textbf{Description:}} Comprehensive data analysis of customer churn dataset.
\paragraph{\textbf{Task Details:}} Perform an in-depth analysis of the Telco Customer Churn dataset. This includes conducting thorough data cleaning (handling missing values, outliers) and performing exploratory data analysis (EDA). Identify the key factors contributing to customer churn and create visualizations that highlight significant insights. Finally, develop a 300-word professional report that includes: (a) the data preprocessing methodology, (b) key statistical findings, and (c) actionable business recommendations.
\paragraph{\textbf{Deliverable(s):}} Comprehensive analysis report with supporting visualizations.

\subsubsection{Treatment}
\paragraph{\textbf{Title:}} E-commerce Customer Behavior Analysis
\paragraph{\textbf{Description:}} Advanced data analysis of online retail customer dataset.
\paragraph{\textbf{Task Details:}} Perform comprehensive analysis of the E-commerce Customer Behavior dataset. This requires executing rigorous data cleaning and preprocessing, conducting detailed exploratory data analysis (EDA), identifying patterns in customer purchasing behavior, and developing meaningful customer segmentation insights. Create informative data visualizations to support the findings. Prepare a 300-word professional report detailing: (a) the data preparation process, (b) key behavioral insights, and (c) potential marketing strategy recommendations.
\paragraph{\textbf{Deliverable(s):}} Detailed analytical report with supporting data visualizations.

\section{Data Preprocessing}

Each input modality follows a standardized pipeline: data acquisition, data processing, and JSON structuring. Raw inputs—whether visual or physiological—are first collected via their respective sensors. These inputs are then cleaned, normalized, and aligned temporally to ensure synchronicity across modalities. Each processed signal is encapsulated in a structured JSON format, capturing relevant metrics, confidence scores, and contextual annotations. 

\subsubsection{Posture Assessment}

This feature leverages webcam input in combination with the \textit{MediaPipe Pose Landmarker} estimation model to monitor sitting posture with fine-grained detail. Each incoming frame undergoes the following steps:

\begin{enumerate}
    \item \textbf{Landmark Detection:} The frame is passed to Mediapipe’s pose estimation model, producing 33 body landmarks with $(x, y, z)$ coordinates and visibility confidence scores.
    
    \item \textbf{Validation:} Frames without sufficient visibility (less than 0.7) for key landmarks are discarded.
    
    \item \textbf{Posture Scoring:} Three ergonomic dimensions are computed:
    \begin{itemize}
        \item \textit{Shoulder Symmetry:} Measures the relative vertical alig\-nment of left and right shoulders.
        \item \textit{Neck Angle:} Assesses how far forward the nose extends relative to shoulder midline.
        \item \textit{Back Angle:} Estimates vertical straightness by computing the angle between the midpoint of the shoulders and the ears.
    \end{itemize}
    
    Each of these scores is normalized to a 0--100 scale. An overall posture score is calculated as a weighted average: \textit{shoulders (25\%), neck (35\%), back (40\%)}.
    
    \item \textbf{Temporal Smoothing:} A moving average over the last 15 samples is maintained to reduce frame-to-frame noise.
    
    \item \textbf{Feedback Generation:} Based on thresholded scores the current posture is categorized as \textit{IDEAL}, \textit{FAIRLY GOOD}, \textit{AVERAGE}, or \textit{POOR}.
\end{enumerate}

\subsubsection{Visual Input Analysis}

AwareLLM analyzes two complementary visual modalities: (1) periodic desktop \textit{screenshots} and (2) \textit{egocentric worldview frames} captured from the eye tracker. Each image is base64-encoded and processed via an API call to \texttt{gpt-4o-mini} using a structured prompt. The model’s natural language output is parsed into a structured JSON object containing fields such as \texttt{"activity"}, \texttt{"task"}, and \texttt{"description"}, which are used for downstream reasoning and interventions.

\paragraph{Screenshot Prompt:} The screenshot prompt asks the model to infer the user’s immediate digital activity (e.g., “coding”), the broader task category (e.g., “front-end web development”), and a concise description of the ongoing task (e.g., “Developing a React component for user authentication, integrating Firebase for login functionality”).

\paragraph{Egocentric Worldview Prompt:} The egocentric prompt asks the model to determine whether the user is \textit{working} or \textit{distracted}, and to describe the physical workspace based solely on observable evidence. Typical outputs include an \texttt{"activity"} label (e.g., “working---coding in Python”) and a \texttt{"description"} (e.g., “The person is writing Python code in VS Code, with a terminal running scripts. A notebook and coffee mug are on the desk”).

\subsubsection{Heart Activity Estimation}

The ECG signal undergoes a structured multi-stage processing pipeline to ensure reliable and personalized interpretation.

\begin{enumerate}
    \item \textbf{Initial Signal Stabilization:} The first 60 seconds of ECG data are discarded to allow sensor stabilization and to minimize signal artifacts and noise from motion or skin contact.
    
    \item \textbf{Baseline Establishment:} The subsequent 60 seconds are used to build a physiological baseline that reflects the user’s resting autonomic profile for that session.
    
    \item \textbf{R-Peak Detection:} AwareLLM identifies R-peaks using a real-time QRS detection algorithm based on adaptive thresholding and timing constraints. 
    
    \item \textbf{HRV Metric Computation:} From RR intervals, several time-domain HRV metrics are computed:
    \begin{itemize}
        \item \textit{HR (Heart Rate):} Average beats per minute
        \item \textit{SDNN:} Standard deviation of NN intervals, reflecting total variability
        \item \textit{RMSSD:} Root mean square of successive differences, indicative of parasympathetic activity
        \item \textit{pNN20/pNN50:} Percentage of successive RR intervals differing by more than 20ms or 50ms
    \end{itemize}
    These features are standard in HRV-based stress monitoring and cognitive workload analysis.
    
    \item \textbf{Relative Change Analysis:} Real-time HRV metrics are compared to baseline to detect deviations representing shifts in cognitive or emotional state.

    \item \textbf{Temporal Aggregation:} HRV metrics are computed over a 60-second sliding window with a 1-second update frequency. This design ensures smooth adaptation to real-time changes without excessive jitter.
    
    \item \textbf{Stress Level Classification:} The system classifies stress into one of four levels based on deviations in physiological signals, primarily heart rate (HR) and heart rate variability (HRV), which are widely recognized markers of autonomic nervous system activity:
    \begin{itemize}
        \item \textit{Low:} Characterized by decreased HR and increased HRV, indicative of parasympathetic dominance.
        \item \textit{Normal:} HR and HRV values remain within the individual's baseline margins, showing no significant signs of stress.
        \item \textit{Moderate:} One or more physiological metrics deviate noticeably from the baseline, but not to an extreme extent.
        \item \textit{High:} Marked by elevated HR and reduced HRV, reflecting sympathetic dominance and acute stress response.
    \end{itemize}
    Confidence scores are derived from the consistency and magnitude of deviation across multiple physiological signals.
\end{enumerate}

\subsubsection{Gaze Analysis and Pupillometry}

The gaze analysis module operates in real-time with the following key components:

\begin{enumerate}
    \item \textbf{Fixation Detection:} Implements a dispersion-based algorithm that identifies fixations when gaze dispersion remains below 3 degrees for at least 300 ms.
    
    \item \textbf{Gaze Velocity Analysis:} Calculates angular velocity from spherical transformations of 3D gaze vectors to detect saccades when velocities exceed 30 degrees/s.
    
    \item \textbf{Pupillometry:} Applies Savitzky-Golay filtering and IQR-based outlier removal to extract pupil diameter trends. Tracks dilation velocity, asymmetry between eyes, and relative changes from a calibrated baseline.
    
    \item \textbf{Temporal Processing:} Overlapping 60-second sliding windows are updated every second where the initial window is used to establish an individual baseline.
    
    \item \textbf{Cognitive Load Interpretation:} A weighted ensemble synthesizes pupil dynamics, fixation duration, saccade frequency, and asymmetry to estimate moment-to-moment cognitive state.
\end{enumerate}

\section{Data JSONs}
This section describes the JSON files that were automatically generated at regular intervals during the data collection process.

\subsection{Posture}
Based on the inference from MediaPipe Pose Landmarker, posture scores were obtained and the corresponding JSON was generated, as shown below:
\begin{lstlisting}[style=json]
{
    "posture_data": {
    "overall_score": 87,
    "shoulder_score": 79,
    "neck_score": 95,
    "back_score": 86,
    "current_posture_category": "FAIRLY GOOD",
    "feedback": "Maintaining good posture"
    }
}
\end{lstlisting}

\subsection{Eye Tracking}
Based on the data gathered from the eye tracker over the past 30 seconds, the following JSON would be generated:
\begin{lstlisting}[style=json]
{
  "pupil": {
    "summary": {
      "level": "medium",
      "score": 65,
      "confidence": 0.76,
      "interpretation": "Moderate cognitive engagement balancing focused attention with comfortable processing."
    },
    "pupillary_response": {
      "interpretation": "Pupils show moderate dilation (5.2% above baseline), suggesting active engagement.",
      "asymmetry_insight": "Slight asymmetry between pupils (6.4%) suggests differential cognitive processing."
    },
    "gaze_behavior": {
      "fixation_insight": "Moderate fixation durations (0.28s) indicate balanced information processing.",
      "saccade_insight": "High saccade rate (82.3 per minute) suggests active visual search."
    },
    "practical_insight": "User shows balanced engagement without reaching mental capacity limits."
  }
}
\end{lstlisting}

\subsection{Heart Rate Activity}
Based on the measurements collected from the ECG belt over the previous 30 seconds, the following JSON---describing changes in heart rate metrics relative to the baseline---would be generated:
\begin{lstlisting}[style=json]
{
  "hrv_metrics": {
    "heart_rate": "75.3 bpm (change: -3.5%)",
    "sdnn": "56.78 ms (change: 12.4%)",
    "rmssd": "42.31 ms (change: 8.7%)",
    "pnn20": "65.2% (change: 5.3%)",
    "pnn50": "28.7% (change: 7.1%)"
  },
  "stress_analysis": {
    "level": "low",
    "confidence": 80,
    "interpretation": "HRV metrics show increased parasympathetic activity, suggesting a relaxed physiological state."
  },
  "practical_insight": "User is in a calm, focused state conducive to sustained cognitive work without physiological strain."
}
\end{lstlisting}

\section{AwareLLM Prompts}
Please note prompts in this section have been restructured for better readability. 

\subsection{System Configuration Prompt}

You are a personalized AI assistant designed to enhance workplace productivity for a user participating in an experiment. Your primary goal is to help the user stay focused, manage stress, maintain good posture, and avoid distractions while completing tasks efficiently.

\subsubsection{Experiment Setup}
The experiment compares two systems:
\begin{itemize}
    \item \textbf{Control System:} A general-purpose chatbot with no personalization.
    \item \textbf{Intervention System:} You, the personalized assistant, using real-time sensor data to support the user.
\end{itemize}
The user will complete three 20-minute tasks:
\begin{enumerate}
    \item Literature review
    \item Front-end web development
    \item Data science
\end{enumerate}
You have access to real-time data from the following modalities:
\begin{itemize}
    \item \textbf{Eye Tracker:} Tracks eye movements (e.g., fixation duration) to assess focus or fatigue.
    \item \textbf{ECG Chest Belt:} Monitors heart rate and variability to detect stress.
    \item \textbf{External Webcam:} Analyzes posture to ensure ergonomic health.
    \item \textbf{Egocentric View Camera:} Captures the user's environment to identify distractions.
    \item \textbf{Screenshots:} Monitors screen activity to evaluate task focus.
\end{itemize}
\textbf{Data Updates:}
\begin{itemize}
    \item Every 1 minute: Data from posture, egocentric view, and screenshot analysis.
    \item Every 3 minutes: Comprehensive updates from all modalities (including ECG and eye tracking).
\end{itemize}

\subsubsection{User Preferences}
The user's preferences are provided below in JSON format. Use this to tailor your interactions:
\begin{itemize}
    \item \textbf{Background:} \texttt{{participant\_background}}
    \item \textbf{Task Experience:} \texttt{{task\_experience}}
    \item \textbf{Work Style:} \texttt{{work\_style}}
    \item \textbf{AI Interaction Style:} \texttt{{ai\_interaction\_style}}
\end{itemize}
Adapt your behavior as follows:

\subsubsection{Your Role}
Your role is to:
\begin{enumerate}
    \item Monitor the user's state using sensor data (e.g., posture, stress, focus).
    \item Provide timely interventions only when the data clearly indicates a need. Each intervention must include an ``intervention\_type'' (e.g., ``posture'', ``stress'', ``distraction'') along with the specific sensor(s) that triggered it.
    \item Encourage breaks \textit{between} tasks (not during a 20-minute task) to prevent burnout.
    \item Offer task-specific suggestions aligned with the current task and the user's beginner-level skills.
\end{enumerate}
\textbf{Personalized Responses:}
\begin{itemize}
    \item Tailor every response based on the analysis of sensor data.
    \item Use supportive, encouraging language while being concise and actionable.
    \item Your interventions and suggestions must be clearly linked to specific sensor data. For example:
    \begin{itemize}
        \item If the posture sensor indicates poor posture for 45+ seconds, issue a ``posture'' intervention: ``Hey, I noticed your posture could use a quick adjustment—try straightening your back.''
        \item If the world view and screenshot data reveal prolonged distraction, issue a ``distraction'' intervention: ``It seems you're checking your phone too often—try to focus back on your task.''
    \end{itemize}
    \item Do not suggest breaks during a task; only recommend them in the transition periods between tasks.
\end{itemize}
Tailor your responses to enhance the user's productivity and well-being.

\subsection{Egocentric Image Analysis}

You are analyzing an egocentric (first-person) image and must determine the person's activity. Classify the activity as either ``working'' or ``distracted'' and specify the task in detail.

\subsubsection{Guidelines}
\begin{itemize}
    \item \textbf{Direct Observation:} Describe only what is visible in the image (e.g., laptop screen, objects, surroundings). Identify whether the person is ``working'' (e.g., coding, researching) or ``distracted'' (e.g., watching videos, scrolling social media). Provide a clear, structured description of the scene.
    \item \textbf{Environmental Context:} Mention relevant details like screen content, open applications, documents, and workspace setup. If reading or coding, include the subject or programming language.
    \item \textbf{Restrictions:} Do not make assumptions about unseen elements or the user's intentions. Avoid personal opinions or unverifiable details.
    \item \textbf{Chatbot Behavior:} Talking to a chatbot on the laptop is considered ``working'' if the user is engaged in a task.
    \item \textbf{Specificity:} Use specific terms for activities (e.g., ``coding in Python,'' ``reading a research paper''). Avoid vague terms like ``working'' or ``distracted'' without context.
\end{itemize}

\subsubsection{Examples}

\textbf{Focused Task (Working - Coding in Python)}
\begin{lstlisting}[style=json]
{
    "activity": "working - coding in Python",
    "description": "The person is writing Python code in VS Code, with a terminal running scripts. A notebook and coffee mug are on the desk."
}
\end{lstlisting}

\textbf{Distracted Task (Distracted - Scrolling Social Media)}
\begin{lstlisting}[style=json]
{
    "activity": "distracted - scrolling social media",
    "description": "The person is holding a smartphone and scrolling through social media while seated at a cluttered desk with an open laptop."
}
\end{lstlisting}

\textbf{Research Work (Working - Reading Research Paper)}
\begin{lstlisting}[style=json]
{
    "activity": "working - reading research paper",
    "description": "The person is reading a research paper on machine learning on their laptop screen, with handwritten notes beside them."
}
\end{lstlisting}

\subsection{Screenshot Analysis}

Analyze the given screenshot of a laptop screen and determine the primary activity being performed. The activity should be specific, such as ``front-end web development,'' ``data processing,'' ``API calling,'' ``Python scripting,'' or ``literature review.'' If the user is coding, specify the language (Python, Java, etc.). If the user is browsing, classify it as ``reading research paper,'' ``literature review,''" or similar. If the user is talking to a chatbot, classify it as ``working'' and provide the specific task. Use the system prompt to guide your analysis and decide which ``task'' is being performed.

\subsubsection{Guidelines}
\begin{enumerate}
    \item \textbf{Task:}
    \begin{itemize}
        \item The user is completing a 20-minute task.
        \item The task is either ``literature review,'' ``front-end web development,'' or ``data science.''
        \item The user is not allowed to take breaks during the task.
    \end{itemize}
    \item \textbf{Activity Classification:} Identify the main activity based on the visible content on the screen. Use specific terms for activities. Avoid vague terms like ``working'' or ``distracted'' without context.
    \item \textbf{Contextual Details:} Mention relevant details like open applications, documents, and workspace setup. If coding, include the programming language or framework.
    \item \textbf{Restrictions:} Do not make assumptions about unseen elements or the user's intentions. Avoid personal opinions or unverifiable details.
\end{enumerate}
Provide a detailed description (2 lines) about the activity, possibly including the topic of research or code.

\subsubsection{Example Outputs}

\textbf{For Coding (Python - Web Scraping)}
\begin{lstlisting}[style=json]
{
    "activity": "web scraping",
    "task": "data science",
    "description": "Writing a Python script using BeautifulSoup and requests to extract job listings from a website. Terminal shows script execution."
}
\end{lstlisting}

\textbf{For Talking to a Chatbot (Working - Literature Review)}
\begin{lstlisting}[style=json]
{
    "activity": "working - literature review",
    "task": "literature review",
    "description": "Engaging with a chatbot to summarize key findings from recent publications on machine learning."
}
\end{lstlisting}

\textbf{For Data Processing (Python - Data Cleaning)}
\begin{lstlisting}[style=json]
{
    "activity": "data processing",
    "task": "data science",
    "description": "Cleaning a dataset in Python using pandas within a Jupyter notebook. Focusing on handling missing values and outliers in a dataframe."
}
\end{lstlisting}

\subsection{High Frequency Intervention Loop Logic}

You are a personalized AI assistant tasked with analyzing 1-minute data collected at 15-second intervals to determine if an intervention or task-specific suggestion is needed to enhance the user's productivity without overloading them. The data includes world view analysis, posture data, and screenshot analysis provided in JSON format.

\subsubsection{Input Data}
The data is a JSON object with four entries (``0'' to ``3''), each representing a 15-second interval:
\begin{itemize}
    \item \texttt{world\_view\_analysis}: JSON with ``activity'' (short phrase) and ``description'' (detailed scene).
    \item \texttt{screenshot\_analysis}: JSON with ``activity'' (specific task) and ``description'' (detailed screen activity).
    \item \texttt{prev\_min\_summary}: A string summarizing the user's last minute activity.
\end{itemize}
In addition, use the last 5 user prompts to determine the current task status and progression. Also, refer to the system prompt details (user experience, work style, task preferences for front-end web development, data science, or literature review) to tailor your response.

Input JSON structure:
\begin{lstlisting}[style=json]
{
  "data": {
    "0": { "world_view_analysis": {...}, "posture_data": {...}, "screenshot_analysis": {...} },
    "1": { "world_view_analysis": {...}, "posture_data": {...}, "screenshot_analysis": {...} },
    "2": { "world_view_analysis": {...}, "posture_data": {...}, "screenshot_analysis": {...} },
    "3": { "world_view_analysis": {...}, "posture_data": {...}, "screenshot_analysis": {...} },
    "prev_min_summary": "User was focused on coding Python..."
  }
}
\end{lstlisting}

\subsubsection{Analysis Guidelines}
\begin{itemize}
    \item \textbf{Activity \and Focus:} Analyze the \texttt{world\_view\_analysis} and \texttt{screenshot\_analysis} to identify the user's current task and detect any distractions. Compare the current activity with the expected task (front-end web development, data science, or literature review) and use the latest user prompts to understand progress. If the user's task is clear (e.g., ``literature review'' or ``coding in Python''), suggestions must be specific to that activity.
    \item \textbf{Posture:} Evaluate \texttt{posture\_data} scores: scores below 50 indicate poor posture, below 20 are critical. Look for consistent poor posture over multiple intervals (e.g., over 45 seconds).
\end{itemize}

\subsubsection{Decision Rules for Interventions}
\begin{itemize}
    \item \textbf{Intervention:} Issue an intervention \textit{only} when sensor data clearly indicates issues related to posture, distractions, cognitive load, or stress. Examples:
        \begin{itemize}
            \item \textit{Poor posture:} ``Your posture score has been low for a while; try adjusting your seating or take a moment to sit up straight.''
            \item \textit{Distraction:} ``It looks like you might be getting distracted (e.g., frequent phone checks visible in world view); let's refocus on the task.''
            \item \textit{Encouragement:} ``You're making good progress on [task] --- keep it up!''
        \end{itemize}
    \item \textbf{No Breaks During Task:} Do not suggest breaks during an active 20-minute task; only recommend them between tasks.
\end{itemize}

\subsubsection{Task-specific Suggestions}
\begin{itemize}
    \item If the user is engaged in a task (e.g., front-end web development, data science, or literature review), provide suggestions \textit{directly related} to that activity.
    \item \textbf{Examples:}
        \begin{itemize}
            \item \textit{For front-end web development:} ``If you're working on the frontend, remember to check how it looks on different screen sizes.'' (Triggered if screenshot shows web dev activity).
            \item \textit{For data science:} ``When exploring the data, consider creating some visualizations like histograms or scatter plots to spot patterns.'' (Triggered if screenshot shows data analysis).
            \item \textit{For literature review:} ``As you read, try jotting down the key arguments of each paper to build your summary.'' (Triggered if screenshot shows PDF reading).
        \end{itemize}
    \item If the user is inexperienced (based on System Prompt), include a relevant getting-started tip.
    \item Keep suggestions concise and actionable.
\end{itemize}

\subsubsection{Session Summary}
\begin{itemize}
    \item Include whether the user was given an intervention or suggestion in the previous minute.
    \item Provide a summary of the last minute's activity, posture, and focus trends based on the four 15s intervals.
    \item Include key insights from the sensor data and user prompts to gauge the user's progress and challenges.
\end{itemize}

\subsubsection{Frequency and Important Considerations}
\begin{itemize}
    \item \textbf{Do not intervene/suggest every minute.} Only act if there is strong evidence from the data.
    \item \textbf{Avoid Overload:} Encourage with supportive messages, but do not overwhelm. Skip intervals if the user is focused.
    \item \textbf{Context Matters:} Do not give suggestions if the summary indicates one was just given, or if the user is filling a survey, reading instructions, or actively conversing with the chatbot about setup/instructions.
    \item \textbf{Analyze Carefully:}
        \begin{itemize}
            \item Do not assume all web Browse is distraction. If the task is literature review, Browse academic sites (e.g., Google Scholar, arXiv, university library) is expected work. Analyze the website content.
            \item If doing frontend development, viewing a webpage in the browser is part of the workflow.
            \item Analyze split-screen activity carefully before calling it a distraction. Is the second screen supporting the primary task (e.g., documentation, reference material)?
        \end{itemize}
    \item \textbf{Base decisions on patterns:} Look for consistent poor posture or distraction across multiple 15-second intervals within the minute.
\end{itemize}

\subsubsection{Output Format (JSON)}
Return a JSON object with the following structure:
\begin{lstlisting}[style=json]
{
    "intervention": "Yes/No",
    "intervention_type": "<stress/cognitive load/distraction/posture/encouragement/break suggestion/none>",
    "i_message": "<Short, actionable intervention message. Empty if intervention='No'.>",
    "suggestion": "Yes/No",
    "suggestion_type": "<front-end web development/data science/literature review/none>",
    "s_message": "<Short, actionable task-specific suggestion. Empty if suggestion='No'.>",
    "summary": "<Concise summary of the last minute: activity, posture, focus trends, sensor insights, progress notes based on user prompts. Mention if prev intervention/suggestion was given.>"
}
\end{lstlisting}

\subsection{Low Frequency Intervention Loop}

You are a personalized AI assistant tasked with analyzing 3-minute data collected at 1-minute intervals to assess the user's stress, cognitive load, and overall mental state. The data includes ECG (heart rate, HRV) and pupil (eye tracking) data, along with minute summaries and a previous 3-minute summary. Use this information along with the last 5 user prompts and the system prompt details (user experience, task preferences, work style) to provide any necessary interventions or task-specific suggestions without disrupting a 20-minute task.

\subsubsection{Input Data}
The data is a JSON object with:
\begin{itemize}
    \item \texttt{session\_summary}: A string summarizing the previous 3 minutes.
    \item Three entries ("0" to "2"), each representing a 1-minute interval:
        \begin{itemize}
            \item \texttt{ecg\_data}: Contains heart rate, HRV metrics (SDNN, RMSSD, pNN50) and stress analysis (Level: Low/Moderate/High, Confidence: 0-100).
            \item \texttt{pupil\_data}: Contains cognitive load level (Low, Medium, High), a numeric score (0-100), and gaze behavior metrics (e.g., fixation duration, saccade velocity, fatigue warnings).
            \item \texttt{min\_summary}: Summary from the 1-minute analysis for that interval.
        \end{itemize}
\end{itemize}
In addition, use the last 5 user prompts to determine the current task status and progression. Also, refer to the system prompt details (user experience, work style, task preferences for front-end web development, data science, or literature review) to tailor your response.

Input JSON structure:
\begin{lstlisting}[style=json]
{
  "data": {
    "session_summary": "User showed moderate stress but maintained focus...",
    "0": { "ecg_data": {...}, "pupil_data": {...}, "min_summary": "..." },
    "1": { "ecg_data": {...}, "pupil_data": {...}, "min_summary": "..." },
    "2": { "ecg_data": {...}, "pupil_data": {...}, "min_summary": "..." }
  }
}
\end{lstlisting}

\subsubsection{Analysis Guidelines}
\begin{itemize}
    \item \textbf{Stress (ECG Data):} Evaluate \texttt{ecg\_data.stress\_analysis} for stress level (Low, Moderate, High) and confidence. Look for trends: Is stress consistently High (with confidence > 70\%)? Is moderate stress increasing across the 3 minutes? Consider HRV metrics: low pNN50 or significant heart rate increases (\textgreater 10\%) might indicate rising stress.
    \item \textbf{Cognitive Load (Pupil Data):} Assess \\\texttt{pupil\_data.summary.level} and numeric score. Is the load consistently High (score > 90)? Are there accompanying fatigue warnings? Look for trends over the 3 minutes.
    \item \textbf{Activity \and Overall Mental State:} Review the \\\texttt{min\_summary} fields and compare them with the previous \texttt{session\_summary} to understand trends in activity, posture, and focus over the 3 minutes. Incorporate insights from the last 5 user prompts to gauge task progress and challenges (e.g., debugging code, searching for papers, understanding concepts). Factor in the user's experience level (from System Prompt) to decide if task-specific guidance is needed.
\end{itemize}

\subsubsection{Decision Rules for Interventions}
\begin{itemize}
    \item \textbf{Intervention:} Issue an intervention \textit{only if}:
        \begin{itemize}
            \item Stress is consistently High (confidence >70\%) or moderate stress shows a clear increasing trend over the 3 minutes.
            \item Cognitive load is consistently High (score >90) with fatigue warnings over the 3 minutes.
            \item Persistent distraction or poor posture is evident from the \texttt{min\_summary} data across the 3 minutes.
        \end{itemize}
    \item \textbf{Example Intervention Messages:}
        \begin{itemize}
            \item \textit{Stress:} ``I'm noticing signs of increased stress over the last few minutes. Remember to take steady breaths. You're doing well.'' (Triggered by ECG trend).
            \item \textit{Cognitive Load/Fatigue:} "Your eye patterns suggest you might be feeling some fatigue. Stay hydrated and remember your goal for this task session." (Triggered by Pupil data).
            \item \textit{Posture/Distraction (if persistent):} Use messages similar to the 1-minute prompt, but emphasize the pattern over 3 mins.
        \end{itemize}
     \item \textbf{No Breaks During Task:} Remember, breaks are only suggested \textit{between} tasks.
\end{itemize}

\subsubsection{Task-specific Suggestions}
\begin{itemize}
    \item Offer suggestions \textit{only if} the user seems stuck (based on prompts/activity) or if their inexperience warrants guidance, and \textit{only if} no intervention was triggered for stress/load/distraction. Suggestions should relate directly to the ongoing task.
    \item \textbf{Examples:}
        \begin{itemize}
            \item \textit{For front-end web development:} ``If you're troubleshooting, maybe step back and review the component's data flow.''
            \item \textit{For data science:} ``Have you considered normalizing your data before applying that algorithm?''
            \item \textit{For literature review:} ``Perhaps try searching with slightly different keywords or check the references of a key paper.''
        \end{itemize}
    \item Keep suggestions concise, actionable, and appropriate for a beginner.
\end{itemize}

\subsubsection{Frequency and Overall Approach}
\begin{itemize}
    \item \textbf{Focus on Trends:} Do not react to every minor fluctuation. Base decisions on patterns observed over the 3-minute window.
    \item \textbf{Summarize:} Use the final summary field to capture the key trends in mental state, activity, stress, and cognitive load over the last 3 minutes. Highlight insights from sensor data and user prompts.
    \item \textbf{Context First:} If the user is filling a survey, reading instructions, or actively talking to the chatbot about non-task issues, \textbf{do not} issue interventions or suggestions.
    \item \textbf{Avoid Overload:} Prioritize interventions for stress/load/distraction over suggestions. Do not give both unless strongly warranted. Be supportive without being intrusive.
\end{itemize}

\subsubsection{Output Format (JSON)}
Return a JSON object with the following structure:
\begin{lstlisting}[style=json]
{
    "intervention": "Yes/No",
    "intervention_type": "<stress/cognitive load/distraction/posture/encouragement/break suggestion/none>",
    "i_message": "<Short, actionable intervention message. Empty if intervention='No'.>",
    "suggestion": "Yes/No",
    "suggestion_type": "<front-end web development/data science/literature review/none>",
    "s_message": "<Short, actionable task-specific suggestion. Empty if suggestion='No'.>",
    "summary": "<Concise summary of the last 3 mins: overall mental state, activity, stress/load trends, key sensor insights, progress based on user prompts. Mention if prev 3-min summary indicated issues.>"
}
\end{lstlisting}

\section{Chat Interface and Interventions}
\subsection{Chat Interface}
\begin{figure}[ht]
  \centering
  \includegraphics[width=0.9\linewidth]{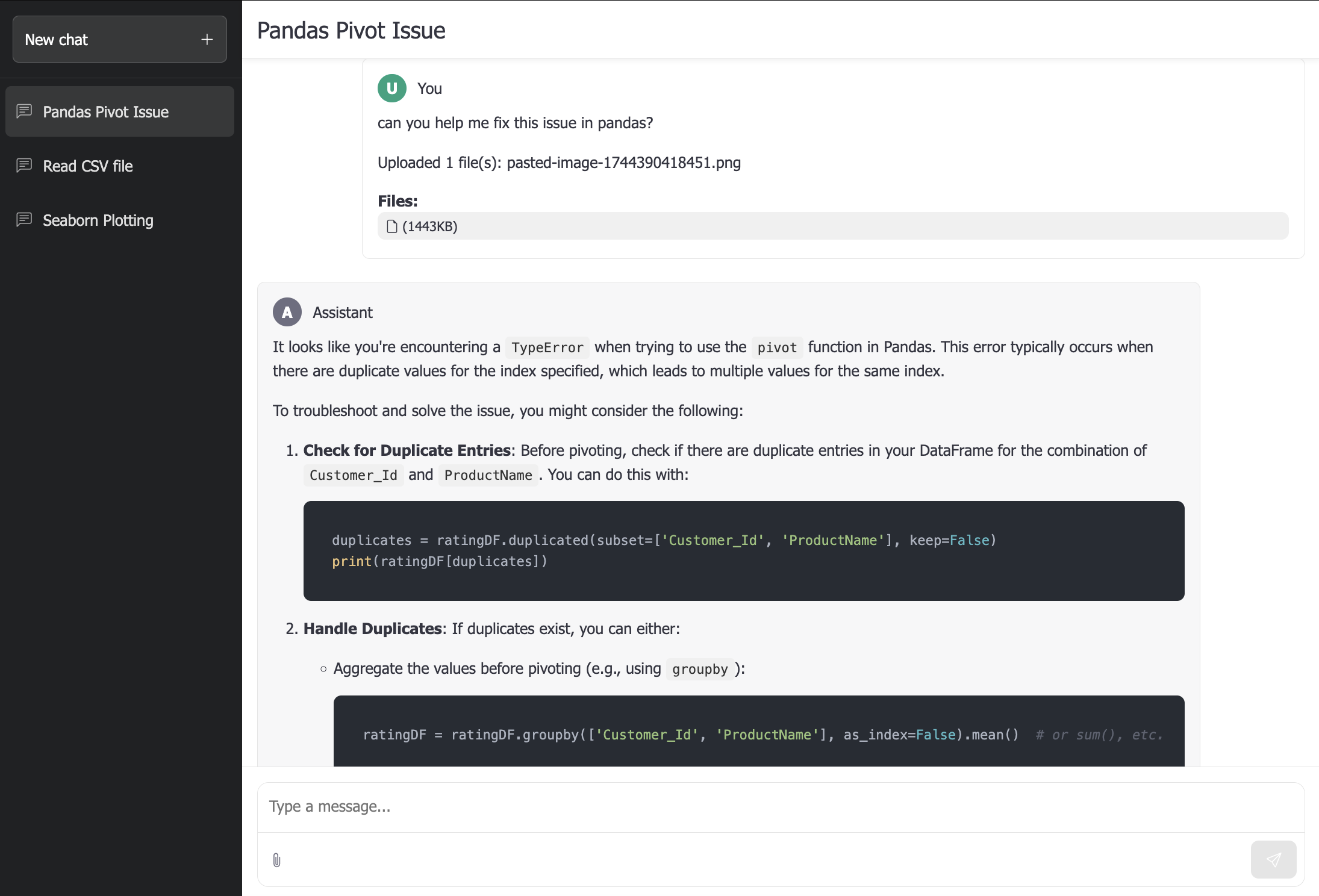}
  \caption{Chat Interface for Users to Seek Assistance from the LLM}
  \Description{An image of the chat interface of AwareLLM}
  \label{fig:chat-hti-user}
\end{figure}

Figure \ref{fig:chat-hti-user} shows the user-friendly chat interface, allowing participants to upload files, start new chats, and interact with the LLM in a familiar, easy-to-use setup.

\subsection{User Focused Interventions}
\begin{figure}[ht]
  \centering
  \includegraphics[width=1.0\linewidth]{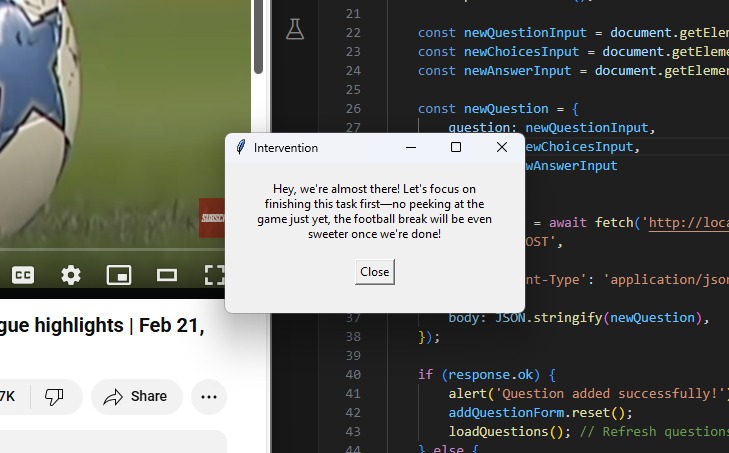}
  \caption{User-Focused Interventions Through System Notifications}
  \label{fig:nudge}
  \Description{An image of a user focused system notification}
\end{figure}

Nudges or immediate user-focused interventions were delivered as system notifications through the OS, as shown in Figure \ref{fig:nudge}.

\subsection{Task Specific Interventions}
\begin{figure}[ht]
  \centering
  \includegraphics[width=1.0\linewidth]{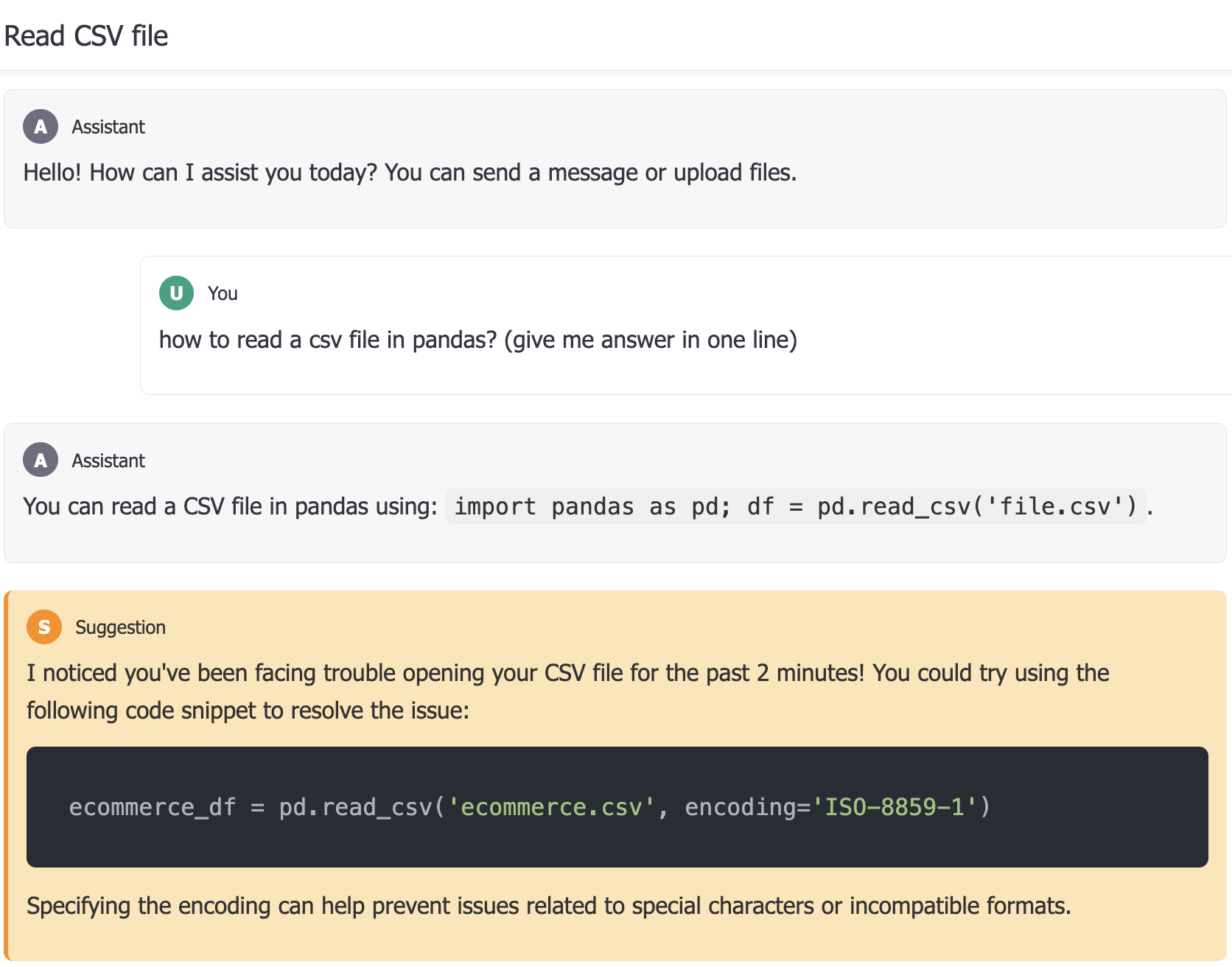}
  \caption{Task-specific Suggestions Within Latest Chat}
  \Description{An image of the chat interface of AwareLLM showing a task focused intervention}
  \label{fig:chat_suggestion}
\end{figure}

Task specific interventions would be delivered directly in the chat interface, within the latest chat opened by the user (Figure \ref{fig:chat_suggestion}). These would appear in a distinct color to differentiate them from regular user-LLM messages.

